\documentclass[final,12pt]{elsarticle}
\journal{Elsevier}

\usepackage{showlabels}
\usepackage{amsfonts}
\usepackage{color}
\usepackage{relsize}
\usepackage{graphicx}
\usepackage{dcolumn}
\usepackage[mathscr]{euscript}
\usepackage{bm}
\usepackage{amssymb,amsmath,amsthm,graphicx,epstopdf,setspace}

\newcommand{\uz}  {\mbox{\boldmath{$u$}}}
\newcommand{\Kz}  {\mbox{\boldmath{$K$}}}
\newcommand{\Oz}  {\mbox{\boldmath{$0$}}}
\newcommand{\Tz}  {\mbox{\boldmath{$T$}}}
\newcommand{\xz}  {\mbox{\boldmath{$x$}}}
\newcommand{\bo} {\widehat{\beta}}
\newcommand{\ta} {\tan\alpha}

\begin{document}
\begin{frontmatter}

\title{
Fractality in selfsimilar minimal mass structures}


\author{ D. De Tommasi$^1$, F. Maddalena$^2$, G. Puglisi$^1$, F. Trentadue$^1$.}
\address{$^{1}$ Dip. Scienze Ingegneria Civile e Architettura, Politecnico di Bari,  Via Re David 200, Bari, Italy}
\address{$^{2}$ Dip.  Meccanica,  Matematica  e Management, Politecnico di Bari,  Via Re David 200, Bari, Italy}




\begin{abstract} In this paper we study
the diffusely observed occurrence of Fractality and Self-organized Criticality in mechanical systems. We analytically show, based on a prototypical compressed tensegrity structure, that these phenomena can be viewed as the result of the contemporary attainment of mass minimization and global stability in elastic systems. \end{abstract}

\end{frontmatter}

\section*{Introduction}

In recent years the experimental evidence of fractal systems has increasingly interested many technological and theoretical fields of research \cite{Man}. Nowadays, Fractality is  recognized as a paradigm of material and structure morphological optimization. Indeed, through billions years evolution, nature developed complex, hierarchical multiscale structures delivering performances unreached by human technologies  \cite{ESL}.
Typical examples of natural hierarchical systems are represented by spider silks \cite{DPSz}, geckoes pads \cite{PTb}, and keratin materials that attain their incredible properties based on the creation of multiscale structure morphologies, often characterized by self-similarity  \cite{IWD}.
The study of the physical mechanisms underlying the diffuse experimental observation of fractal systems is important understanding fundamental features of many nature and biological phenomena and is also crucial for the design of new efficient bioinspired materials and structures.

In this paper, we are interested in an explicit theoretical understanding of  why nature shapes materials and structures in self-similar systems. In a previous paper  \cite{DMPT}, 
 we analyzed the possibility of mass optimization of a compressed bar, by a T-bar tensegrity (see Fig.\ref{Fig1}) with increasing complexity. Differently from the present paper, there we considered only a partial substitution of the compressed bars with self-similar T-bars. Here we extend this approach, by analyzing a full substitution of all  struts and find that in the limit  of increasing slenderness or decreasing load, the structure approaches a fractal type structure. Such kind of scaling relations,  leading to a simultaneous buckling at all scales of a hierarchical structure, have been observed and analytically studied in different contexts, such as gecko pads adhering systems studied in  \cite{PTb}.

To this end we adopt a very simple prototypical, tensegrity type device, aimed at the transmission of compression loads. As we analytically show the requirements of both mass optimization and  global stability   deliver a fractal like morphology characterized by power laws behavior. 
More precisely, starting from the elementary Euler column, we deduce that geometrical complexification by means of self-similar tensegrity structures favors the mass minimization whence the global stability condition is imposed. As a result the 
limit structure exhibits fractal dimension, power law behavior, and a contemporary attainment of critical states at all involved scales.  The analytical evidence of the proposed results clarifies
that mass minimization and global stability may be fundamental mechanisms for interpreting the observation of scale-free geometries  in many mechanical systems.

Our prototypical scheme is based on the concept of  {\it tensegrity}, first introduced by Snelson  and afterwards by Fuller \cite{Fu}, \cite{Motro_2003}. These   structures consist of compressed members (struts) connected by tensile cables \cite{C1}. Their main characteristics  are lightness and developability, making them a promising system both in  Civil  \cite{SF} and Aerospace  \cite{DMP} Engineering. 
Also tensegrities  have been recognized as a diffusely adopted way  to transmit and control forces in biological systems (see {\it e.g.} cell cytoskeleton \cite{IWD, Volokh}).The counterpart of the lightness property of these structures is the onset   of instability effects   both at the single struts and global scales  (see \cite{NS},  \cite{C2} and  \cite{CW2}).

The paper is organized as follows. 
In Section~\ref{MRT} we outline the main results of the paper based  
on some simplifying assumptions. More detailed analytical treatment is presented the subsequent sections where some of the simplifying assumptions are removed. In particular we show how  the optimal complexity linearly depends on the logarithm of the (non-dimensional) assigned load. 
 Similarly we obtain a log-log relation between the optimal mass and the assigned load.
 Based on these results, we investigate the fractal character in the limit of decreasing load (or increasing slenderness of the compressed structure) and analytically obtain that the limit $D$ of the structure dimension is not integer. In particular we analytically study the dependence of $D$ on geometrical and constitutive parameters.  Of course, as usual in the context of fractal systems  this (`infinite') mass optimization and refinement, leading to a lower fractal dimension,  is ideal, {\it but clarifies why so many systems in nature exhibit self-similar character}. 
The readers interested in the only main physical results can focus only on  this section.

The  mathematical details of our results are contained in the subsequent sections.
Specifically, in Section~\ref{lsm} we determine the optimal mass of the tensegrity, at the generic complexity, by only considering the  local stability condition.
We study the positive definiteness of the total potential energy of the system and we obtain that the optimal mass condition together with the constraint of global stability delivers the contemporary attainment of instability at all the scales. 
In Section~\ref{opts} we discuss the optimality condition for assigned load, geometric and material parameters.  Then we focus on the analysis of the power laws and fractal limit system obtained for vanishing load. We employ the well known  technique of box counting \cite{Falconer} to estimate the dimension of the region occupied by the optimal  tensegrity. In the Appendix we give all analytical details regarding  the important global stability effect (often neglected in the literature of tensegrity systems).

\section{Main results: optimality of fractal tensegrities}\label{MRT}

We search for an optimal mass structure carrying a given compressive load $P$ by comparing the classical `continuum' column choice with self-similar tensegrity type structures.
The main novelty of our approach is the  explicit solution  (differently from many approaches in the literature: see \cite{SO}) of the {\it global stability} problem for a class of  self-similar structures. The importance  of global stability,  especially in the analysis of optimal complexity and mass minimization approaches, has been recently pointed out in \cite{DMPT}. A global stability analysis approach had then be considered in the more recent paper \cite{SKSK}.   Here, we extend the results in \cite{DMPT} by considering more general complexification of the tensegrity structure and by determining analytical results on the fundamental case of fractal limit systems.

Beside imposing global stability whose analysis is detailed in the Appendix,  we also  assume that the internal axial  forces $N$, $T$ satisfy 
 the optimality design conditions
\begin{equation}\label{optoptopt}
\left \{ \begin{array}{ll}
N=\min \left\lbrace  N_{E},\sigma_y A\right\rbrace & \mbox{ for struts,} \vspace{0.2 cm}\\
T=\sigma_y A & \mbox{ for cables,} \end{array} \right . \end{equation}
where $N_{E}$ is the critical Eulerian axial force, $\sigma_y$ is the limit  value for the axial stress and $A$ is the cross section area of the members. Here for simplicity we consider a material with the same limit traction and compressive stress, but the model can be easily generalized.

To obtain explicit analytic results we focus on the class of plane tensegrities represented in 
Fig.~\ref{Fig1}, with arbitrary complexity, fixed total length $L$, under a given compressive load $P$.  Numerical extensions to the
three-dimensional case has been recently proposed in \cite{DMPTb}.

\subsection{Single compressed bar}
\begin{figure}[h!]\vspace{-0.2 cm}
\centering\includegraphics[height=13.5cm]{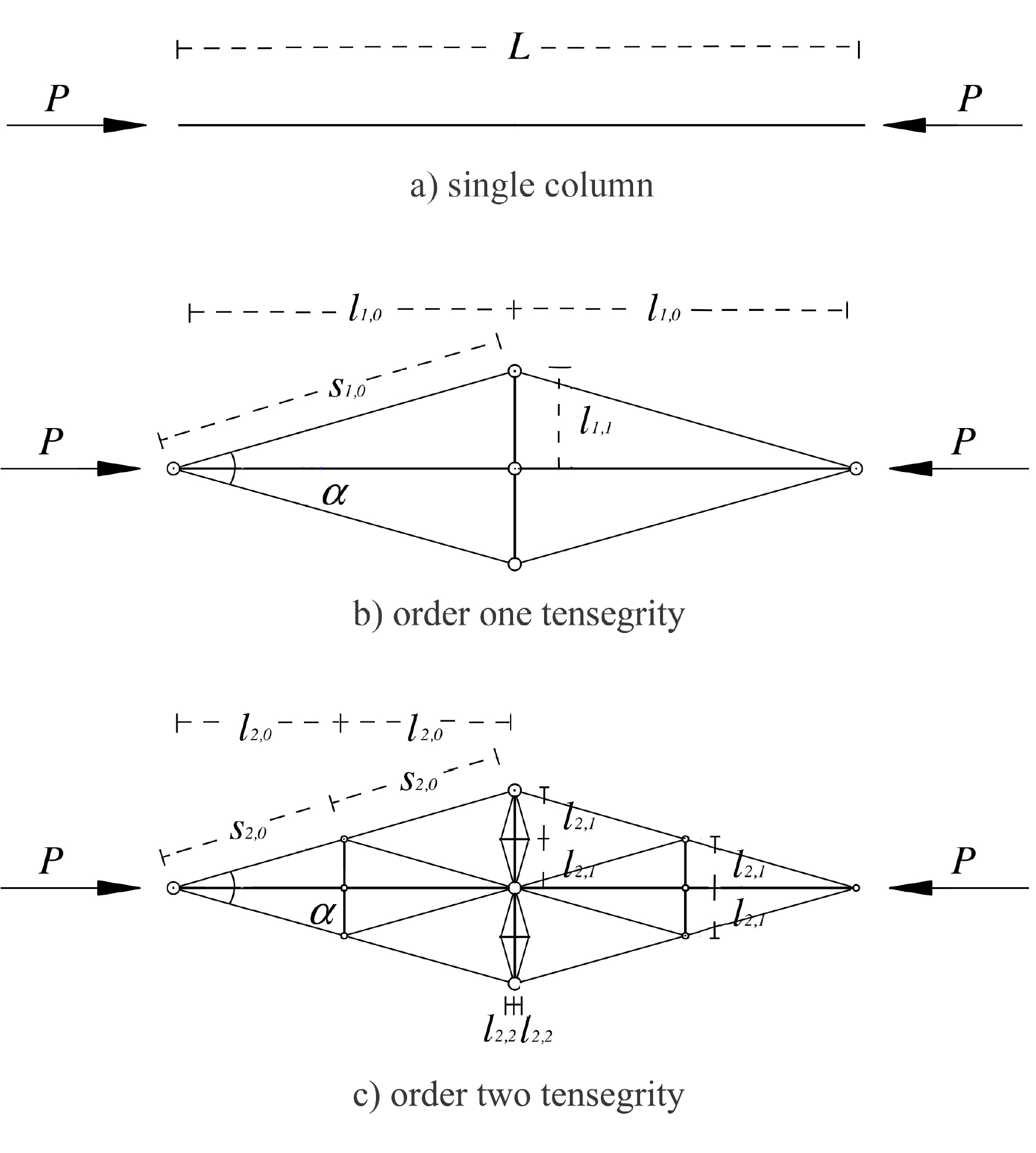}\vspace{-0.3 cm}\caption{Scheme of the self-similar tensegrity column.}
\label{Fig1}
\end{figure}

Consider first a  single  bar (see Fig.~\ref{Fig1}a)   with
cross section of area $A$ and axial area moment $J=\frac{A^2}{\xi^2\pi^2}$ (here $\xi$  is a geometrical non dimensional parameter depending on the cross section shape, {\it e.g.} $\xi^2=\frac{4}{\pi}$ in the case of circular section). 


Observe that, given a compression load $P$, the optimal mass 
is the maximum  between the mass  $m_y$  corresponding to the  material failure and the mass  $m_b$ corresponding to Eulerian buckling, namely 
 
\begin{equation}
\label{masse1}
\begin{array}{l}
m_{y}(L,P)= \rho\frac{P}{E \varepsilon_y} L,\vspace{0.2 cm} \\
m_{b}(L,P)= \rho\, \xi\, L^2\, \sqrt{\frac{P}{ E}}, \end{array} 
\end{equation}
where $ \rho$ is the mass density, $E$ is the Young modulus and 
$\epsilon_y= \frac{\sigma_{y}}{E}$ is the limit strain (e.g. $\epsilon_y  \backsim10^{-3}$ for steel).

In the case  $m_y>m_b$ the result is trivial, because, as we will show, the optimal mass is attained by the single column. Thus we  assume that the failure is attained 
due to buckling ($m_b\geq m_y$) and this implies
\begin{equation}
\sqrt{P}\leq \xi \, L \, \sqrt{E} \,\varepsilon_y .
\label{chid}
\end{equation} 
Let us denote by
\begin{equation}
\chi=\chi(L,P)=\dfrac{1}{\xi \, L \varepsilon_y } \sqrt{\frac{P}{ E}}
\label{pchi}
\end{equation}
  the ratio between the Eulerian critical load $P$ and the compressive failure load 
 $\sigma_y\, A$ of this column.
Thus  \eqref{chid} corresponds to $\chi\leq 1$.
We anticipate that $\chi$  is the main non-dimensional parameter governing the optimal complexity problem studied in this paper. 

\subsection{Tensegrity columns with increasing complexity}

Let us consider the possibility of replacing a single column by a self-similar tensegrity supporting  the same load $P$ at the same distance $L$ with lower mass. 
Specifically we consider the class of plane tensegrities   represented in Fig. \ref{Fig1}, with all  possible  complexities. Here we extend the analysis in \cite{DMPT} where the authors considered the possibility of substituting the only horizontal struts with tensegrities. Of course the special choice of planar tensegrity in Fig.\ref{Fig1} is related to our aim of attaining analytical results keeping as clear as possible the physical meaning of our results and in particular on the relation between self similarity, stability and mass minimization, fundamental both in many physical problems and in the field of new materials design.
To the knowledge of the authors, there exists no general theory regarding such connections, and that no rigorous (analytical, variational) deduction of fractal mass minimizer are available.  In this regards we refer to the recent paper \cite{DMPTb}.
 
 We begin by a tensegrity consisting, in the terminology of \cite{SO}, in a T-bar  (see 
 Fig.~\ref{Fig1}b)  characterized by the same length $L$.
 All struts are  assumed hinged and have the same shape of the cross-section  (fixed $\xi$) of the zero order column and the same material parameters $E$ and $\sigma_y$. 
 Here we measure the prestress by
   \begin{equation}
   \beta =\frac{N_{1}}{P}>1,
   \label{beta_def}
   \end{equation} 
 where $N_{1}$ is the compressive force in the two horizontal struts of length $L/2$.
 For assigned prestress the other compression and tension internal forces can be easily deduced by equilibrium considerations. 
 
 The higher order tensegrities are obtained by substituting all struts by self-similar structures. In the structure of order $two$ we assume that the horizontal struts of length $L/4$ carry an axial force $N_2=\beta^2 P$.
  Finally at the complexity $n$ the horizontal struts of length $L/2^n$ undergo a compressive force   
	\begin{equation}N_n=\beta^n P.
	\label{Pn}
	\end{equation}
  
In the Appendix we prove that the stiffness tangent matrix is positive definite if and only if $\beta>\widehat\beta$, where
 \begin{equation}
 \widehat{\beta}:=\frac{\varepsilon_y+\tan^2 \alpha}{\tan^2 \alpha \left(1-\varepsilon_y\right)}\label{BB}.
 \end{equation}
  In Section~\ref{opts} we show   that  {\it the optimal mass condition (see \eqref{bet}), under the 
global stability constraint,  leads to the contemporary attainment of buckling at all the scales}. Such condition is reached when $\beta=\widehat\beta$, corresponding to semi-definite positiveness of the tangent stiffness matrix.
  This represents one of the main results of this paper.

It is worth noticing that even if, for the sake of simplicity in this paper we have considered  a unique parameter $\beta$ governing the prestress,  in \cite{DMPT} the authors showed that a scale invariant prestress can be deduced by mass minimization and stability results.  
 
Roughly speaking (see Section \ref{optoptoptopt}), {\it refinement (increasing complexity) leads to mass optimization until material failure is attained} (indeed, in view of Eq.~\eqref{Pn}, normal forces increase with the complexity $n$).
In this section, to focus on the main physical results of the paper we assume that under refinement the material failure is first attained in the struts of the load horizontal  axis. This situation is attained when the condition \eqref{qbqbqb}, regulated by geometrical and constitutive parameters, is fulfilled.
 The alternative  case in which material failure is first attained in smallest struts  will be anyway considered in the following sections.
 
Under this hypothesis, the optimal mass is obtained by simply comparing at each complexity order $n$ the buckling and resistance mass of the $2^n$ horizontal struts of length $L/2^n$ under the load $\widehat\beta^n P$. 
Explicitly we have
\begin{equation}
\label{masse10}\begin{array}{l}
m_{y}(L/2^n, \widehat\beta^nP)= \rho\frac{\widehat\beta^n P}{\sigma_y} \frac{L}{2^n},\vspace{0.2 cm} \\
m_{b}(L/2^n, \widehat\beta^n P)= \rho\, \xi\, (\frac{L}{2^n})^2\, \sqrt{\frac{\widehat\beta^n P}{ E}}. \end{array} \end{equation}
Hence at the complexity $n$, the condition \eqref{chid} becomes 
\begin{equation}
\chi \leq \left (2 \sqrt{\widehat\beta}\right )^{-n}.
\label{PP}
\end{equation}
Observe that in Section \ref{optoptoptopt} it will be shown that, when the  condition \eqref{qbqbqb} is not assumed, that \eqref{PP} admits the   following generalization
\begin{equation}\chi \leq \left (\dfrac{2 \sqrt{\widehat\beta}}{q}\right )^{-n},
\label{PPP}\end{equation} 
where
\begin{equation}
q=\min\lbrace {\bar q,1 \rbrace},
\label{q0}\end{equation}
\noindent with
\begin{equation}\label{qbar}
\bar q= \sqrt{\frac{\widehat\beta\tan\alpha}{\widehat \beta-1}}.\end{equation}

The inequality  \eqref{PPP} can be rephrased as  
 \begin{equation}
n\leq c_f:=\frac{\log\bigl( \frac{1}{\chi}\bigr)}{ \log \bigl( \frac{2\sqrt{\widehat\beta}}{q}\bigr)  },
\label{chif00}
\end{equation}
 so that  {\it \eqref{chif00}  determines the values $n\in \mathbb N$ of the complexity 
  for which  the  failure in all struts is due to buckling.}
 

 Under the condition \eqref{chif00},
in Sect \ref{optoptoptopt} we obtain that the  optimal mass, for any natural $n$, can be expressed as follows 
\begin{equation}
M_n=M^{s}_n+M^{b}_n=m_b(L,P) p^n+ m_s(L,P) \left (t^n-1 \right ),   \label{totalmass}
\end{equation}
where 
\begin{equation}\label{ppp}p=\frac{\sqrt{\bo}+\sqrt{(\bo-1)\tan^5\alpha} }{2}\end{equation} and 
\begin{equation}\label{ttt} t=(\bo -1 )\tan\alpha^2 +\bo>1.\end{equation}
By using  \eqref{totalmass}-\eqref{ttt} we get that if the following inequalities are fulfilled:
\begin{equation} \left\lbrace  \begin{array}{lll}
p&<&1\\
n&<&c_b:=\dfrac{\log \chi - \log\left( \dfrac{t(1-p)}{p(t-1)}\right)}{\log\left( \dfrac{p}{t}\right)}
\end{array}\right.
\label{c_b}
\end{equation}
then  the following  monotonicity condition holds true
\begin{equation}\label{francomaddalena}
M_{n} - M_{n-1}<0.\end{equation}
In  order to detect the optimal complexity   $\widehat{n}$ 
we have to compare the  inequalities \eqref{chif00} and \eqref{c_b}. 
To this aim we  observe that 
\begin{equation}
\frac{c_b}{c_f}= \frac{\log 2 \sqrt{\widehat\beta}}{\log \frac{t}{p}} \left (1 +\frac{\log \frac{t(1-p)}{p(t-1)}}{\log\chi }\right ),\end{equation}
and,  in the limit of vanishing $\chi$ of interest in this paper, we have
\begin{equation}
\lim_{\chi\rightarrow 0}\frac{c_b}{c_f}=  \frac{\log 2 \sqrt{\widehat\beta}}{\log \frac{t}{p}}>1.
\end{equation}
Thus $c_b>c_f$ definitively as $\chi\rightarrow 0$ and  {\it the optimal complexity $\widehat{n}$ is greater or equal than $c_f$}. Moreover, by \eqref{chif00} it is easy to recognize that
\begin{equation}
\label{chin}
c_f \rightarrow \infty\:\:\:\:\hbox{\rm as}\:\:\:\:  \chi\rightarrow 0
\:\:\Rightarrow\:\: \widehat{n}\rightarrow \infty. 
\end{equation}
  
   In the following,  to obtain explicit solutions,  we will fix 
$\widehat n =c_f $. In this regards we point out that the mass \eqref{totalmass} evaluated at 
$\widehat n=c_f $
represents an upper bound for the \textit{exact} optimal mass.  In view of   \eqref{chif00} we note that the complexity grows as the (non-dimensional) load $\chi$   decreases
\begin{equation}\hat n =-\frac{2\log \chi}{\log (\frac{4\widehat \beta}{q})}.
\label{nopt}\end{equation}

\begin{figure}[h]
\centering $$\begin{array}{ll}
\includegraphics[height=4.8cm]{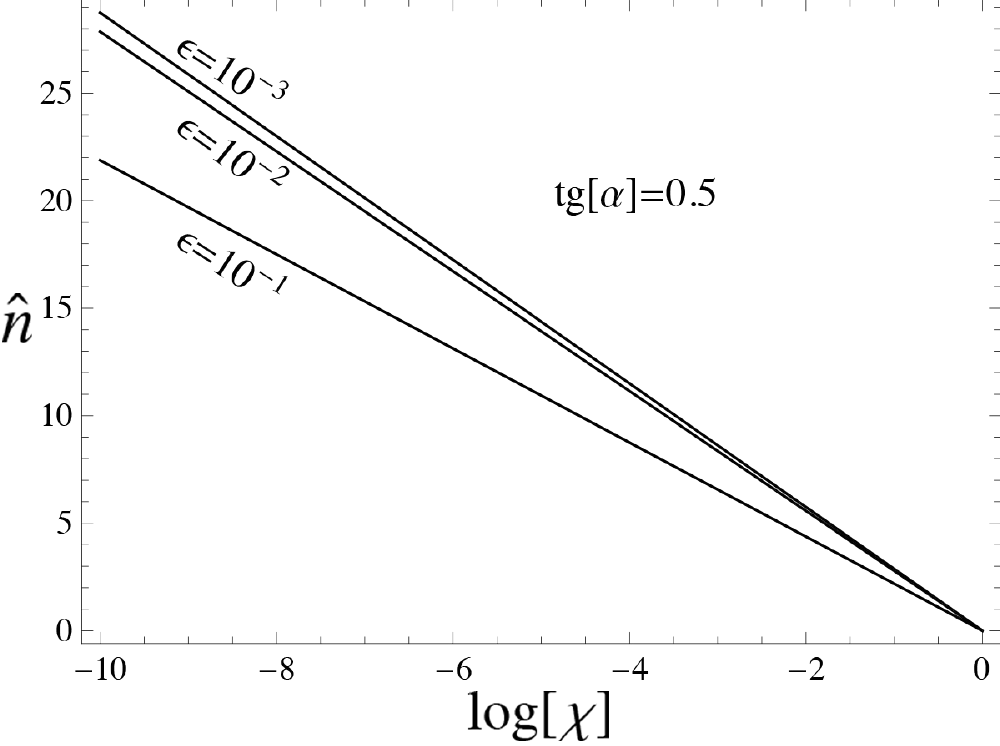}&
\includegraphics[height=4.8cm]{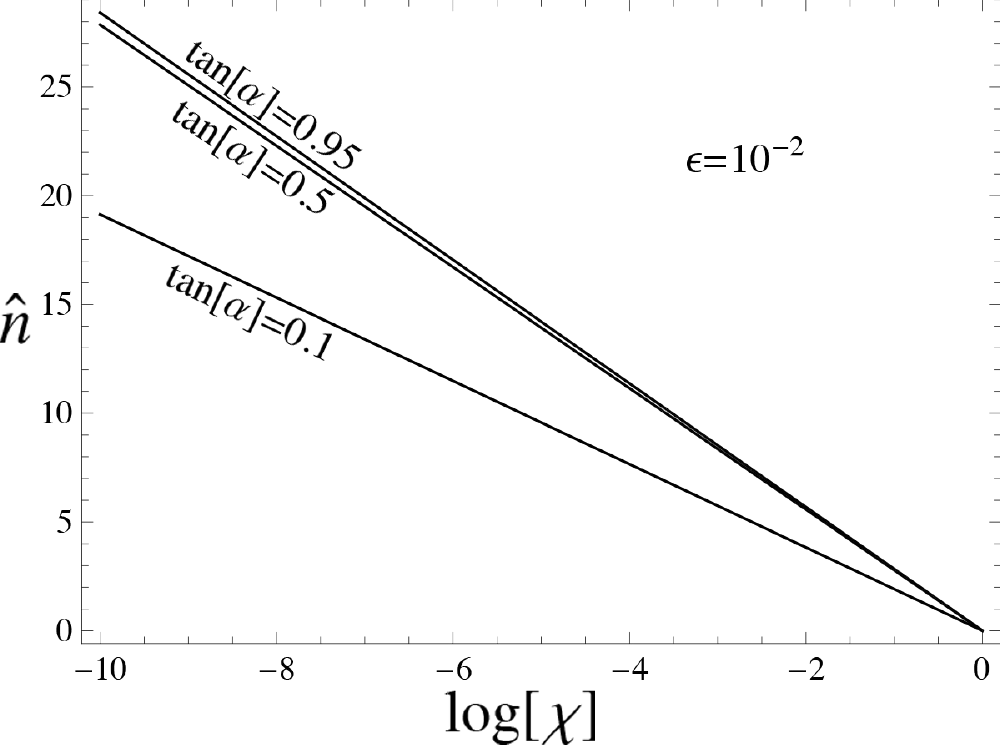}
\end{array}$$
\vspace{-0.5 cm}\caption{Logarithmic dependence of the complexity from the non-dimensional load $\chi$.}\label{Figgg}
\end{figure}

The logarithmic dependence of the complexity is represented in Fig.~\ref{Figgg} showing how the complexity grows as the load is decreased. Observe also that the optimal complexity grows as the limit strain $\epsilon$ decreases and the angle $\alpha$ increases.

\begin{figure}[h!]
\hspace{-1 cm}$$\begin{array}{ll}
\includegraphics[height=4.3cm]{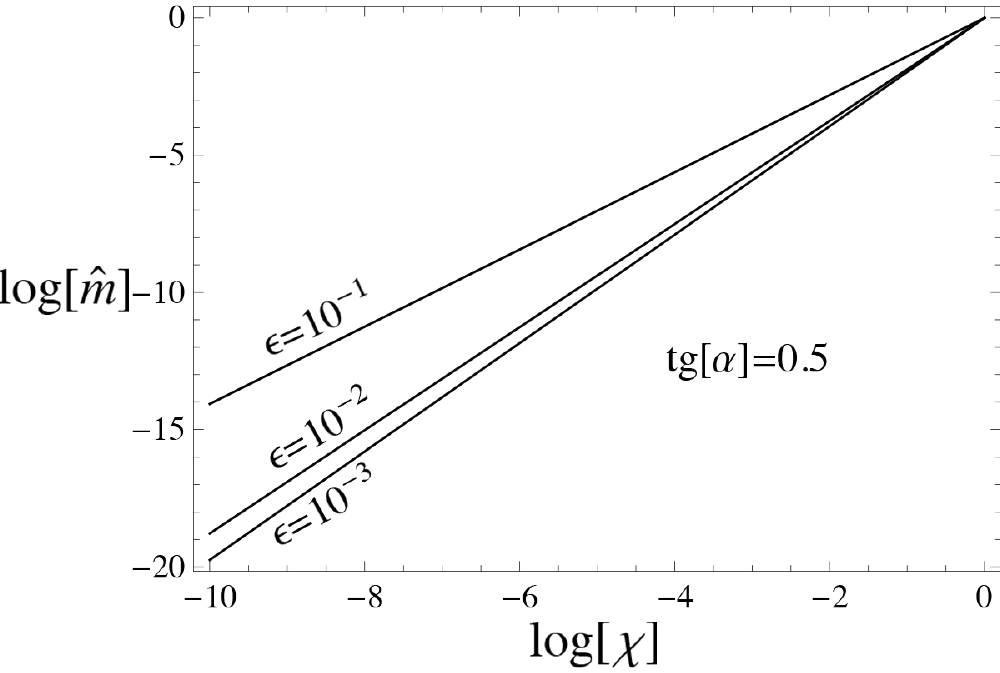}&
\includegraphics[height=4.3cm]{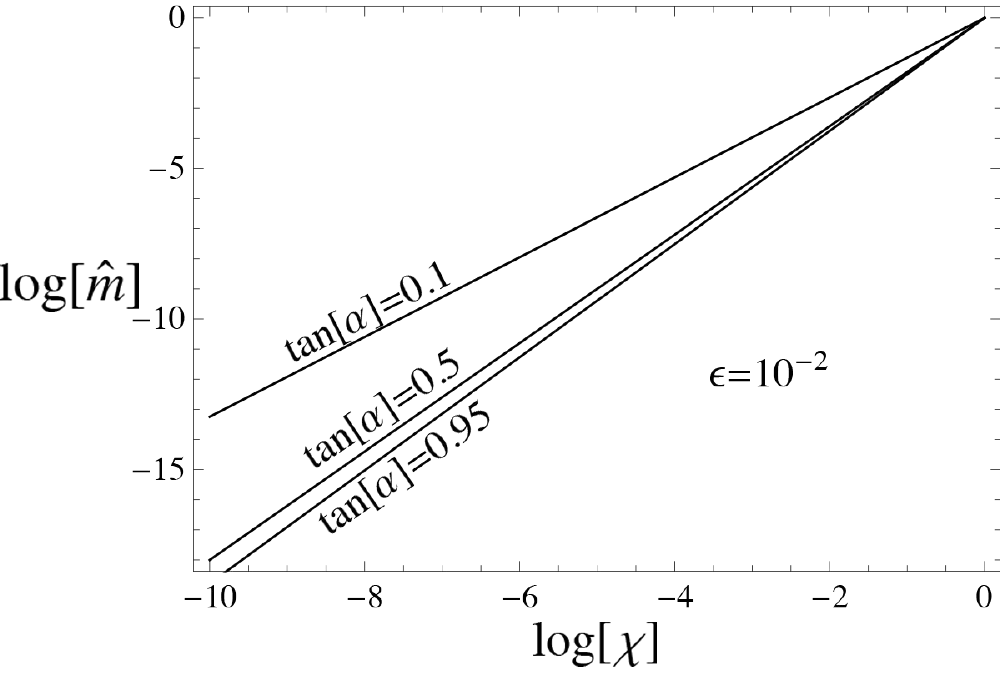}
\end{array}
$$
\vspace{-0.5 cm}\caption{log-log relation between the optimal mass and the non-dimensional load $\chi$.}
\label{Fig2}
\end{figure}

To describe the power law relation between the optimal mass and the dimensionless load $\chi$, we introduce the dimensionless mass
$m_n:=\frac{M_n}{\rho \xi^2 L^3 \varepsilon}$,
whose optimal value $\widehat m$, by \eqref{totalmass}, \eqref{nopt} takes the form
(see Section \ref{optoptoptopt}, Eqs. \eqref{muns} and \eqref{munb})
\begin{equation}
\widehat{m}= \chi\,\, p ^{c_f(\chi)}+\chi^2 \left ( t^{c_f(\chi)}-1 \right ).\label{mmoo}
\end{equation}
{\it The logarithmic law  in Fig.\ref{Fig2}  shows the self-similar scale-independent behavior of the optimal system}. 

Observe that a  linear dependence of the logarithms is attained for small values of the dimensionless load. Indeed, we may rewrite  \eqref{mmoo}
as
\begin{equation}
\widehat{m}= \chi\,p ^{c_f(\chi)}(1+o(\chi )),\label{mmoobis}
\end{equation}
assuming 
\begin{equation}\frac{q\, t}{2p\sqrt{\beta}}<1.\end{equation}
One can verify (see Fig.\ref{figc}) that the last inequality holds true
except in the very special case $\bar q=1$ (see \eqref{qbar}). As a result, as $\chi \rightarrow 0$
we obtain 
 \begin{equation}
\log \widehat m=\left( 1 - \frac{2\log p}{\log \frac{4 \widehat\beta}{q^2}}  \right )\log \chi.
\end{equation}
This equation delivers the analytical value of the critical exponent as a function of the parameters $\alpha$ and $\epsilon$.

\emph{The  optimal mass tensegrity corresponds to a contemporary attainment of instability at all scales.
}\vspace{0.2 cm}

This could  suggest, regarding a more general dynamical setting, 
a  possible analogy to the frequent occurrence of contemporary instabilities at all scales observed in physical phenomena as in the framework of Self Organized Criticality \cite{SOC}.\bigskip

\begin{figure}[h!]\centering$$\begin{array}{ll}
\includegraphics[height=5cm]{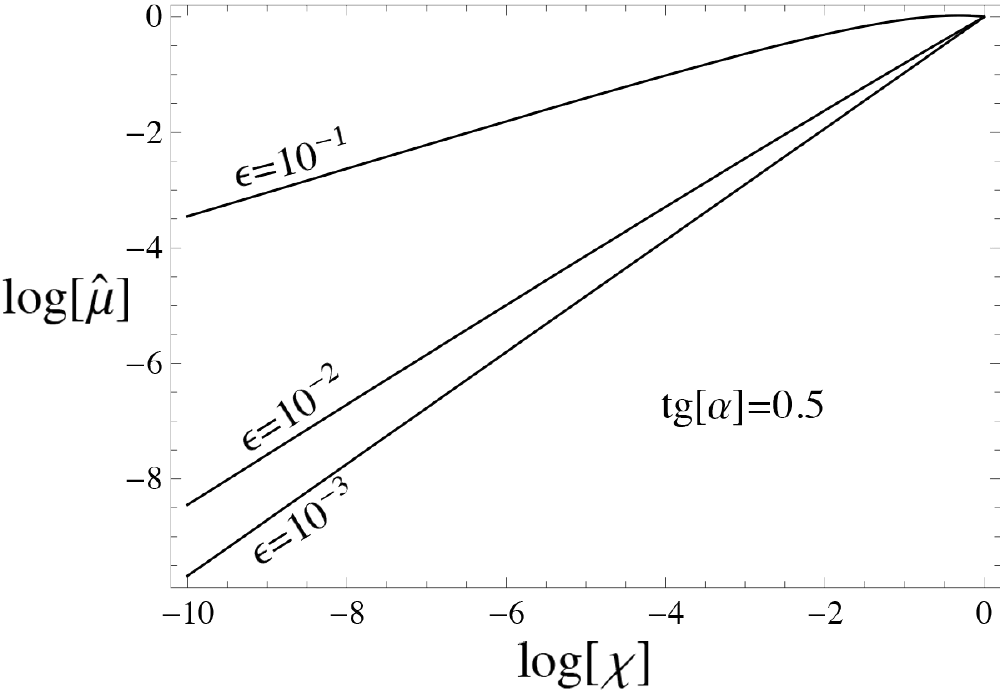}&
\includegraphics[height=5cm]{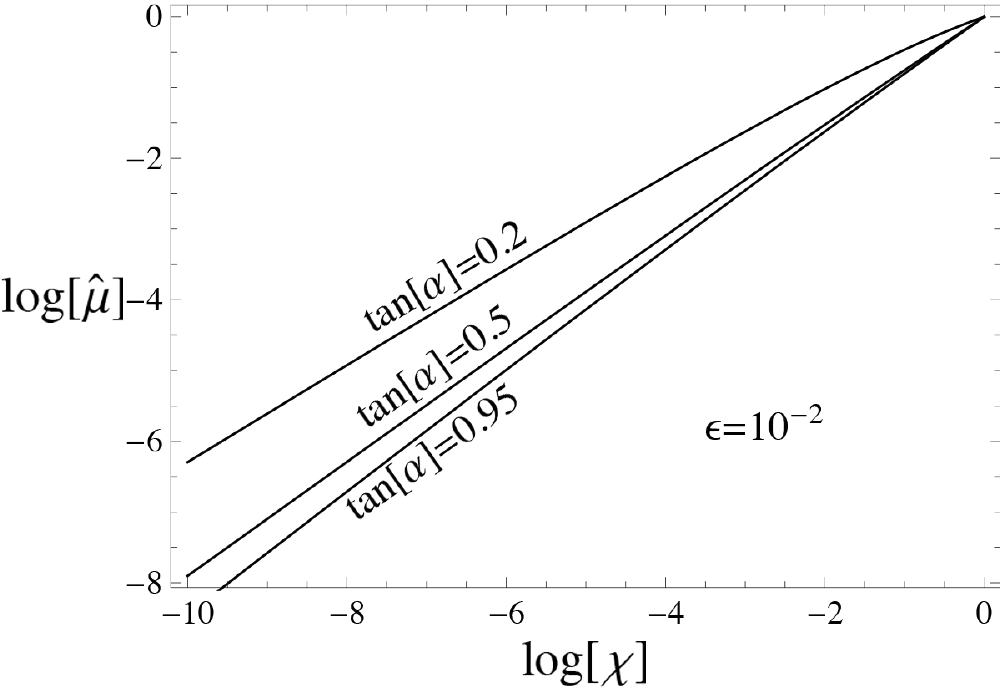}
\end{array}
$$
\vspace{-0.5 cm}\caption{log-log relation between the optimal mass ratio $\bar \mu$ and the non-dimensional load $\chi$.}
\label{Fig3}
\end{figure}

\noindent Remark. {\footnotesize To evidence  the significant mass reduction of the tensegrity here proposed with respect to a single bar,  we consider
the following non-dimensional efficiency parameter
\begin{equation}\mu_n=\frac{M_n}{m_b},\end{equation}
representing the ratio between the tensegrity mass and the mass of the optimal single column supporting the same load.
The optimal value $\widehat\mu$  of $\mu$ is attained at the complexity $c_f$, 
hence 
\begin{equation}
\widehat\mu= p^{c_f(\chi)}+\chi \left (t^{c_f(\chi)}-1 \right ) .
 \label{totalmassad}\end{equation}
The log-log relation between the mass ratio $\widehat\mu$ and the non dimensional load $\chi$  is represented in Fig.\ref{Fig3}. Observe that the efficiency grows as $\chi$ decreases. Moreover the mass gain can be increased also by increasing the angle $\alpha$ and decreasing the limit strain $\epsilon$.
Once again  an exact linear dependence of the logarithms is attained in the limit of vanishing load $\chi$ with
\begin{equation}
\log \widehat\mu=-2\left( \frac{\log p}{\log 4 \widehat\beta}  \right ) \log \chi.
\end{equation}}

\subsection{Tensegrity columns with infinite complexity and fractal limit dimension}

Now
we study the remarkable limit case of  
$\chi\rightarrow 0$ (or analogously $P\rightarrow 0$ or $L\rightarrow\infty$) corresponding,  according with \eqref{chin}, to optimal mass limit systems with infinite self-similar refinement. 
   In the following we  show that  the dimension of the region occupied by the   \textit{optimal} tensegrity tends to a fractal limit $D<3$. 
  
   More precisely,  according to  \cite{Falconer} (Section 3.1), 
    the {\em box-counting dimension} of any subset $\Omega\subset \mathbb R^3$ is  defined  as
   
   \begin{equation}
   \label{dimgen}
   {\rm dim}_B\,\Omega =\displaystyle\lim_{\delta \to 0} \dfrac{\log(N(\delta))}{-\log(\delta)},
   \end{equation}
   where  $N(\delta)$ is the number of elements in  the smallest set of cubes of (non-dimensional) 
	side $\delta$ that covers the region $\Omega$.
   Let    $\Omega_{\widehat n}$ be the region occupied by the tensegrity of complexity $\widehat n$,    
we will show that 
  
  \begin{equation}
  D:=\lim_{\widehat n\rightarrow \infty}\text{dim}_B \,\Omega_{\widehat n}<3.
  \end{equation} 
  
\begin{figure}[b!]
\centering
\includegraphics[height=5cm]{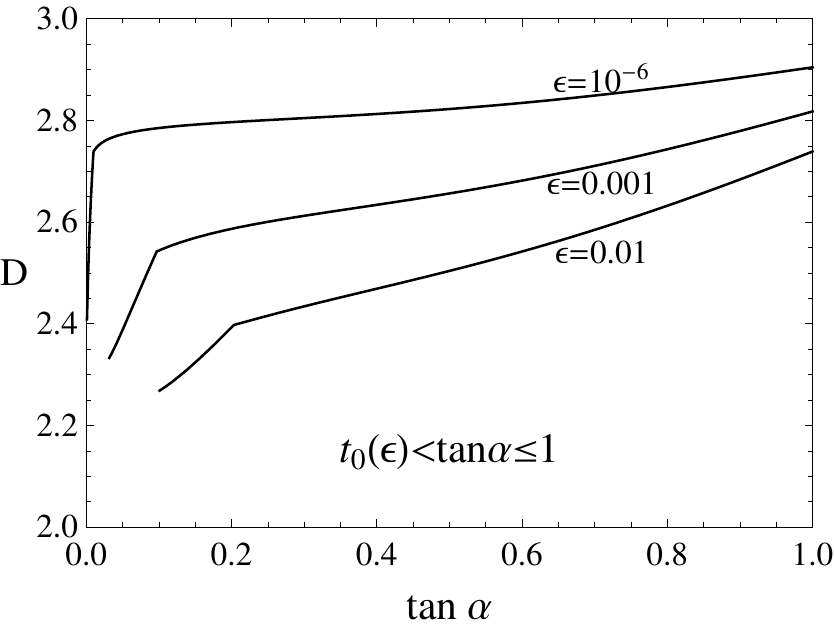}
\vspace{-0.25 cm}
\caption{Limit of the tensegrity dimension as a function of the limit strain $\varepsilon$ 
and of the angle $\alpha$.}
\label{Fig10}
\end{figure}

More precisely, in Section \ref{opts}, considering that  at any order of complexity all the members are slender cylinders ({\it i.e.} the diameters of their cross sections are always much smaller than their lengths) we prove

\begin{equation}
D=
 \dfrac{  \log \left( 2 (1+\frac{\bar{q}}{\tan^3\alpha})\left(  \frac{\bar{q}}{q} \right) ^{3\eta-1} \right)}{{\log \left( \frac{2}{\tan\alpha} \left(  \frac{\bar{q}}{q} \right) ^\eta \right)}}\,,
\end{equation}    
where
\begin{equation}
\eta=\left\lbrace \begin{array}{lll}
1&\mbox{if}& \quad  \bar{q}\ge1\vspace{0.15cm}\\
\frac{1}{2}&\mbox{if}&\quad   \bar{q}<1.
\end{array}\right. 
\end{equation}

\noindent In Fig.~\ref{Fig10} we represent the dependence of the limit dimension from the two parameters $\epsilon$ and $\alpha$.
 Notice that $D$ is decreasing with respect to 
 $\varepsilon$ and one can check that $D$ is bounded below in the range here considered $(t_0(\varepsilon)\leq \tan\alpha\leq 1)$. 

The possibility of evaluating analytically the critical index of the power laws (the fractal type limit system and its dependence from both geometrical and constitutive parameters) is crucial in the perspective of determining the optimum mass properties of light structures and it represents in our opinion a fundamental step in the perspective of interpreting the observation of many hierarchical structures observed in mechanobiology. In particular, we recall that one of the main successful models in cell mechanics is based on the tensegrity concept and also in this case hierarchical morphologies are observed  (see {\it e.g} \cite{IWD} and references therein).

\section{Local stability}\label{lsm}

In this Section we deduce the properties on the minimum mass tensegrities based on the only local stability analysis. The study of global stability in Section \ref{stab} will then determine the optimal mass structures.  We begin by considering the tensegrity of complexity $n=1$ and then we extend the analysis to the cases of tensegrities with increasing complexity represented in Fig.\ref{Fig1}.
 Notice that the analysis of the single column and of the complexity $n=1$ is analogous to the one  in \cite{DMPT}.
We report it for readers convenience. The analysis of higher complexities is instead different due to the choice of substitution of all struts as compared with the previous work.

\subsection{Order 1 tensegrity column}
 To extend to higher complexity, each member and each internal force  is  identified by two indexes  whose meaning will be explained in the following subsection.
The structure of complexity $n=1$ is constituted by\vspace{0.3 cm}
  
\noindent --  two  horizontal struts  of length $ l_{1,1}=\frac{L}{2}$, carrying  the  compressive force  $N_{1,1}=\beta P$, where $\beta$ is the prestress parameter introduced in \eqref{beta_def};\vspace{0.15 cm}

 \noindent --  four  prestressed cables of length $s_{1,0}=\frac{1}{2 \cos \alpha}L $. By equilibrium considerations we find that  the traction  in the four cables is $T_{1,0}=\frac{( \beta -1)}{(2 \cos \alpha)}P$; \vspace{0.15 cm}
 
 \noindent -- two vertical struts  of length $ l_{1,0}= \frac{L\ta}{2}$, carryng the compressive force  $N_{1,0}=( \beta -1)P \tan \alpha $, where $\alpha$ is  the angle between the cables and the horizontal struts.\vspace{0.3 cm}

 The optimal mass of a generic strut with  length $l=aL$, carrying an axial force $N=b P$, is obtained by considering both  the case in which $N$ corresponds to its Eulerian critical load  and  the case  in which $N$ corresponds to its compression failure load. In the first case we determine the minimum buckling (non-dimensional) mass
\begin{equation}\mu_b=\mu_b(a, b)= \frac{m_{b}(aL,bP)}{m_{b}(L,P)}=a^2 \sqrt{b},\label{mub}\end{equation}
whereas in the second case, using (\ref{masse1}),  the minimum material failure mass is given by
\begin{equation}\mu_y=\mu_y(a, b)= \frac{m_{y}(aL,bP)}{m_{b}(L,P)}= \frac{m_{y}(L,P)}{m_{b}(L,P)} \frac{m_{y}(aL,bP)}{m_{y}(L,P)}=\chi \, ab.\label{mus}\end{equation} Thus  the optimal  mass is 
\begin{equation}\mu=\mu(a,b)=\max \{\mu_b(a,b);\mu_y(a,b)\}=\max \{a^2 \sqrt{b};\chi\, a b\}.\label{mu}\end{equation} 
Moreover the minimum mass  of a cable with length $l=aL$,
carrying a traction force $T=bP$, is assigned by
$ \mu_y(a,b)$ and of course, the non-dimensional mass (with respect to $m_b(L, P)$) of the order one tensegrity is the sum of the non-dimensional masses of all its members.

\subsection{Higher Order  tensegrity columns}\label{optoptoptopt}
The tensegrity column of order $2$ is obtained (see Fig.\ref{Fig1}c) by replacing in the  order $1$ tensegrity each  strut by  a geometrically similar tensegrity of order 1,  carrying the same compressive force.
 By reiterating this process  we obtain the $n$-th order  tensegrity. At the generic step $i$ $(0< i\leq n)$ the complexity is increased from $i-1$ to $i$  by replacing each strut with a new T-bar  of the same length,  carrying the same force.   Since each new T-bar is composed by $4$ struts and $4$ cable,  at the step $i$  new $4^i$  struts are  introduced and $4^{i-1}$ struts are removed. Further, $4^i$  cables are added to the preexisting  $4^1+.....+4^{i-1}$ cables. 
 
 We assume that in all T-bars the prestress state is described  by the same parameter $\beta$, so that at the complexity $n$  all struts  and cables of equal length carry equal axial forces. We remark that in this generation process each strut is replaced by two  half length struts having the same direction and carrying an axial force increased by a factor $\beta$ and by two struts, whose length is scaled by a factor  $\frac{\ta}{2}$, orthogonal to the original strut and carrying axial forces scaled by a factor $(\beta-1)\ta$. 
 
 We can consider a partition of  the set of struts in subsets denoted by two indexes:  the first one is the complexity $n$; the second index $i$ denotes the number of steps in the above described generation process  in which the  generated strut  has not changed its direction. In other words   the direction of the strus has been changed with respect to the removed strut (from horizontal to  vertical and \textit{vice versa}) $n-i$  times. Thus the subset  $(n, i)$ contains 
 $2^{n}\binom{n}{i}$  bars of lengths $ l_{n,i}=a_{n,i}L$, carrying the compression forces $N_{n,i}=b_{n,i}P$, where
\begin{equation}\left\lbrace  \begin{array}{lll}
a_{n,i} &=&\Bigl ( \dfrac{1}{2^{i}}\Bigr ) \Bigl ( \dfrac{\ta}{2}\Bigr ) ^{n-i}=\dfrac{\ta^{n-i}}{2^n}\vspace{0.2cm}\\
b_{n,i}&=& \beta^{i}((\beta-1)\ta)^{n-i}
\end{array} \right.\quad  i=0,1,....,n.
\label{cmp1}\end{equation}

 In the same way,  the subsets of cables  having the same length and carrying the same traction force can be identified by two indexes $j$ and $k$.  In particular $j$ $(1\leq j\leq n)$ denotes the step at which the subset has been introduced; the second index $k$ $(0\leq k\leq j-1)$ denotes the  strut $l_{j-1,k}$ replaced by the new T-bar.  Then, in the tensegrity of complexity $n$ we have
 \[4\sum\limits_{j=1}^{n} 2^{j-1}\sum\limits_{k=0}^{j-1}\binom{j-1}{k}\]
cables of length  $s_{j,k}=c_{j,k} L$, carrying  the traction force $T_{j,k}=d_{j,k}P$, where: 
\begin{equation}\label{cccc}
c_{j,k}=\dfrac{1}{2\cos\alpha}\dfrac{l_{j-1,k}}{L}
           =\dfrac{1}{2\cos\alpha} \dfrac{(\ta)^{j-k-1}}{2^{j-1}}
           =\dfrac{(\tan\alpha)^{j-k}}{2^{j} \sin \alpha}\end{equation}
and           
\begin{equation}\label{dddd}        
d_{j,k}= \dfrac{\beta-1}{2\cos\alpha}\dfrac{N_{j-1,k}}{P}= \dfrac{\beta-1}{2\cos\alpha}\beta^{k}(\, (\beta-1)\ta\,) ^{j-1-k}= \dfrac{\beta^{k}(\, (\beta-1)\ta\,) ^{j-k}}{2 \sin\alpha}
\end{equation}
$j=1,....,n, \,\, k= 0,1,....,j-1$.\vspace{0.5 cm}

We are now in position to evaluate  the relative mass  of the order n tensegrity $\mu_n$, which can be written as 
\begin{equation}
\mu_n=\mu_n^s+\mu_n^b,
\label{mun}
 \end{equation}
where $\mu_n^s$ is the total mass of the cables and  $\mu_n^b$ is the total (relative) mass of the bars. In particular, in view of   \eqref{mus}, the contribution $\mu^{s}_n$ is given by
\begin{equation} \begin{array}{lll}
\mu^{s}_n&=&4  \sum\limits_{j=1}^{n}2^{j-1}\sum\limits_{k=0}^{j-1}\binom{j-1}{k}\,\mu_y (c_{j,k}, d_{j,k})\\
&= &4\chi  \sum\limits_{j=1}^{n}2^{j-1}\sum\limits_{k=0}^{j-1} \binom{j-1}{k}\dfrac{(\tan\alpha)^{j-k}}{2^j \sin \alpha} \dfrac{((\beta-1)\tan\alpha)^{j-k}\beta^{k}}{2 \sin \alpha} \vspace{0.1 cm}\\
&=&\chi (\beta -1)(\tan^2\alpha+1)   \sum\limits_{j=1}^{n}\left[\beta+(\beta-1)\tan^2\alpha\right]^{j-1} \vspace{0.1 cm}\\
&=&\chi (\beta -1)(\tan^2\alpha+1) \dfrac{((\beta -1 )\tan\alpha^2 +\beta)^n-1}{(\beta -1)(\tan^2\alpha+1)} \vspace{0.1 cm} \\
&=&\chi(t^n-1),
\end{array}
\label{muns} \end{equation} 
where  \begin{equation}t=(\beta -1 )\tan\alpha^2 +\beta>1.\label{tttt}\end{equation}

Regarding the optimal mass of the struts we have to consider both buckling and failure  by extending the approach in \cite{DMPT}. As a result one can determine the optimal complexity at assigned load $P$. In this paper we are interested in the limit of vanishing load and we now show that in this case the failure of the struts is always due to local buckling.  

To this end, in view of \eqref{chid} and \eqref{pchi}, we  note  that at the complexity $n$ every group $l_{n,i}$ of struts  undergo elastic buckling if and only if
\begin{equation}
\chi( l_{n,i},N_{n,i})=\frac{1}{\epsilon_y \xi\,l_{n,i}}\sqrt{\dfrac{N_{n,i}}{E}}
               =\left( \frac{\sqrt{b_{n,i}}}{a_{n,i}}\right) \dfrac{1}{\epsilon_y \xi\,L}\sqrt{\dfrac{P}{E}}\leq 1\quad i=0,1,...,n.
\label{chil1}\end{equation}
So that,  by  \eqref{cmp1} and \eqref{pchi}, we have   buckling failure in all struts if
\begin{equation}\begin{array}{lll}
\dfrac{1}{\epsilon_y \xi\,L}\sqrt{\dfrac{P}{E}}&\leq &\displaystyle\min_{i=0,1,...,n} { \left\lbrace \frac{a_{n,i}}{\sqrt{b_{n,i}}}\right\rbrace }\vspace{0.15cm} \\
 &=&\displaystyle\min_{i=0,1,...,n} { \left\lbrace \frac{1}{ \left(  2\sqrt{ \beta }  \right) ^n} \left(\sqrt{\frac{\beta\tan\alpha}{\beta-1}}\right)^{n-i}\right\rbrace} \vspace{0.15cm} \\
 &=&\left( \frac{q}{ 2\sqrt{ \beta }}  \right)^n,
 \end{array}
\label{chif0}\end{equation}
where 
\begin{equation}\left\lbrace \begin{array}{lll}
q&=&\min{\left\lbrace  \bar{q},1\right\rbrace}\vspace{0.15cm}\\
\bar q&:=& \sqrt{\frac{\beta\tan\alpha}{\beta-1}}\\
\end{array}\right.  .
\label{qqq}
\end{equation}
Notice that \eqref{chif0}  can be equivalently written as
\begin{equation}
n\leq c_f:=\dfrac{\log(1/ \chi)}{\log\left( \frac{ 2\sqrt{ \beta }}{q}  \right)}.
\label{cf}\end{equation}
That is, if $n\leq c_f$  the failure of bars occurs only for buckling. Since $\lim_{\chi\rightarrow 0}c_f=+\infty$ we obtain the announced result.

A further simplification is obtained if we note that if 
 \begin{equation}\label{qbqbqb}\bar q=\sqrt{\frac{\beta\tan\alpha}{\beta-1}}>1\end{equation}the first group of struts undergoing material failure (with index $i=n$) is constituted by the  struts laying on the horizontal axis of the tensegrity. {\it This result is at the base of the simplified treatment of Section\ref{MRT}}. 
 
 If instead $\bar{q}<1$ the first group undergoing material failure has index $i=0$, that in the case here considered ($\alpha\leq\pi/4$) is constituted by the struts with the smallest length. In the special case $\bar{q}=1$ all  struts simultaneously undergo  material failure.
 
Now, under the assumption $n\leq c_f$, in view of  \eqref{mub}, the  relative  mass of the struts $\mu^{b}_n$ is given by
\begin{equation} \begin{array}{lll}\mu^{b}_n
&=& 2^n \sum\limits_{i=0}^{n}\binom{n}{i}\, \mu (a_{n,i}, b_{n,i}) \\
&=& 2^n\sum\limits_{i=0}^{n}
\binom{n}{i}  \dfrac{\tan\alpha^{2(n-i)}}{2^{2n}}\sqrt{\left( (\beta-1)\tan\alpha\right) ^{n-i}\beta^{i}} =p^n
\end{array}
\label{munb} \end{equation}
where
\begin{equation}p=\dfrac{\sqrt{\beta}+\sqrt{(\beta-1)\tan^5\alpha} }{2}.\label{pppp}\end{equation}

\section{Optimization and fractal limit}
\label{opts} 
 Now, our aim is to minimize the relative mass $\mu_{n}$  of the tensegrity under the global stability constraint. 
Then, assuming \eqref{mun} and \eqref{eq2}, we study the problem
\begin{equation}
 \displaystyle \min \left\lbrace  \mu_{n}( \beta )\:\vert\: \Kz\succeq\bf 0 \right\rbrace .
\label{optprob}
\end{equation}
For any fixed $n\in \mathbb N$, 
in view of \eqref{bet}, the optimization problem \eqref{optprob} reduces to 
\begin{equation}
\displaystyle\min  \left\lbrace \mu_{n}( \beta ) \:\vert\:\beta\geq\widehat{\beta}\right\rbrace .
\label{optprob1}
\end{equation}
Since, by \eqref{mun}, \eqref{muns}, \eqref{munb},   $\mu_{n}$ is  an increasing function of the prestrecthing parameter $\beta$,  the minimum in  \eqref{optprob1} is attained in correspondence of $\beta=\bo$.
{\it The optimum is then characterized by  the contemporary attainment of  instability at all the scales}. Moreover it is possible to show that introducing multiple independent prestress parameters, does not allow any further  mass reduction. 
Therefore the adoption of a single prestress parameter, while greatly simplifying the formulation, does not restrict our result. 

The solution of \eqref{optprob1} depends through \eqref{bet} on the geometric parameter $\tan\alpha$ and the (small) material parameter $\epsilon_y$. For the sake of simplicity, we limit our analysis to the following (realistic for typical values of $\epsilon$) range of the geometric parameter $\tan\alpha$:
\begin{equation}
\sqrt{\dfrac{\epsilon}{1-2\epsilon}}=t_{0}(\epsilon) \leq\tan\alpha\leq 1,
\label{tlim}
\end{equation}
 corresponding, using \eqref{bet}, to the following range of the optimal prestress:
\begin{equation}
\dfrac{\epsilon+1}{1-\epsilon}\leq \widehat{\beta}\leq 2.
\label{blim}
\end{equation}

Regarding the optimization with respect to the complexity $n$ we
first recall that under  the condition \eqref{tlim} we have that $p<1$ (see \eqref{ppp}), as numerically shown in Fig.\ref{figc}. This is a necessary condition to have a decreasing mass \eqref{mun} with increasing complexity $n$.

Moreover in this section we study the important limit case of vanishing external load, namely 
 $\chi\rightarrow 0$ (or equivalently  $L\rightarrow\infty$).  Hence, in view of 
   \eqref{chif00}, we have to study the limit case  $n\rightarrow\infty$.

Finally notice that the stability analysis developed in the previous Section \ref{stab} is based on the hypothesis that all  struts undergo buckling before material failure, and this holds until $n\leq c_{f}$. Here we limit our analysis to the 
\textit{suboptimal} complexity $\widehat{n}=c_{f}$. Of course, since we will show the fractality
of the suboptimal solution, this result can be extended to the, possibly lower, optimal mass.

\subsection{Limit relative mass}
  
 First we observe that in the limit of infinite complexity, by \eqref{munb} and \eqref{muns},  we have that the non dimensional mass $\mu$ attains the limit value

\begin{equation}\begin{array}{ll}
\widehat{\mu}_{\infty}&=\displaystyle \lim_{n \to \infty} \mu_{n}\left( \,\widehat \chi (n)\,\right) \\
&=\displaystyle\lim_{n \to \infty}\left( \: p^{n}+ \left( \frac{q\,t}{ 2\sqrt{ \widehat \beta }}  \right)^{n}-\left( \frac{q}{ 2\sqrt{ \widehat \beta }}  \right)^{n}  \:\right) \vspace{0.15cm}\\
&=\displaystyle \lim_{n \to \infty} p^{n} \left( 1+ \left( \frac{q\,t}{ 2p\sqrt{ \widehat\beta  }}  \right)^{n}- \left( \frac{q}{ 2p \sqrt{ \widehat \beta   }}  \right)^{n} \:\right),
\end{array}
\label{limmu}
\end{equation}
where, $\widehat \chi$ is the inverse function of $n \mapsto c_f (\chi)$ in \eqref{chif00}.

To study \eqref{limmu}, we are going to prove   that  the following inequality holds true:
\begin{equation}
\frac{q\,t}{ 2p\sqrt{ \widehat \beta }}\leq 1,
\label{a31}
\end{equation}
(notice that, since $t>1$, the previous inequality implies $\dfrac{q}{ 2p\sqrt{\widehat \beta }} <1$). 
Indeed, in view of  \eqref{pppp} and \eqref{tttt}, Eq.  \eqref{a31} is equivalent to
 \begin{equation}
 \begin{array}{ll}
 &\sqrt{\widehat \beta (\widehat \beta-1)\tan^5\alpha}\geq q(\widehat \beta-1)\tan^2\alpha+\widehat\beta (q-1).
  \end{array}
  \label{a32}
  \end{equation} 
Since, by \eqref{qqq}, $\widehat\beta (q-1)\leq 0 $, then the above condition \eqref{a32} is satisfied if 
\begin{equation}
\begin{array}{ll}
 & \dfrac{\beta\tan\alpha}{ (\beta-1)}>q^2.
 \label{a33} \end{array} 
 \end{equation} 
Finally, by using again   \eqref{qqq}, we get that \eqref{a33} is  always satisfied.\\
  
\begin{figure}
\centering 
\includegraphics[height=5cm]{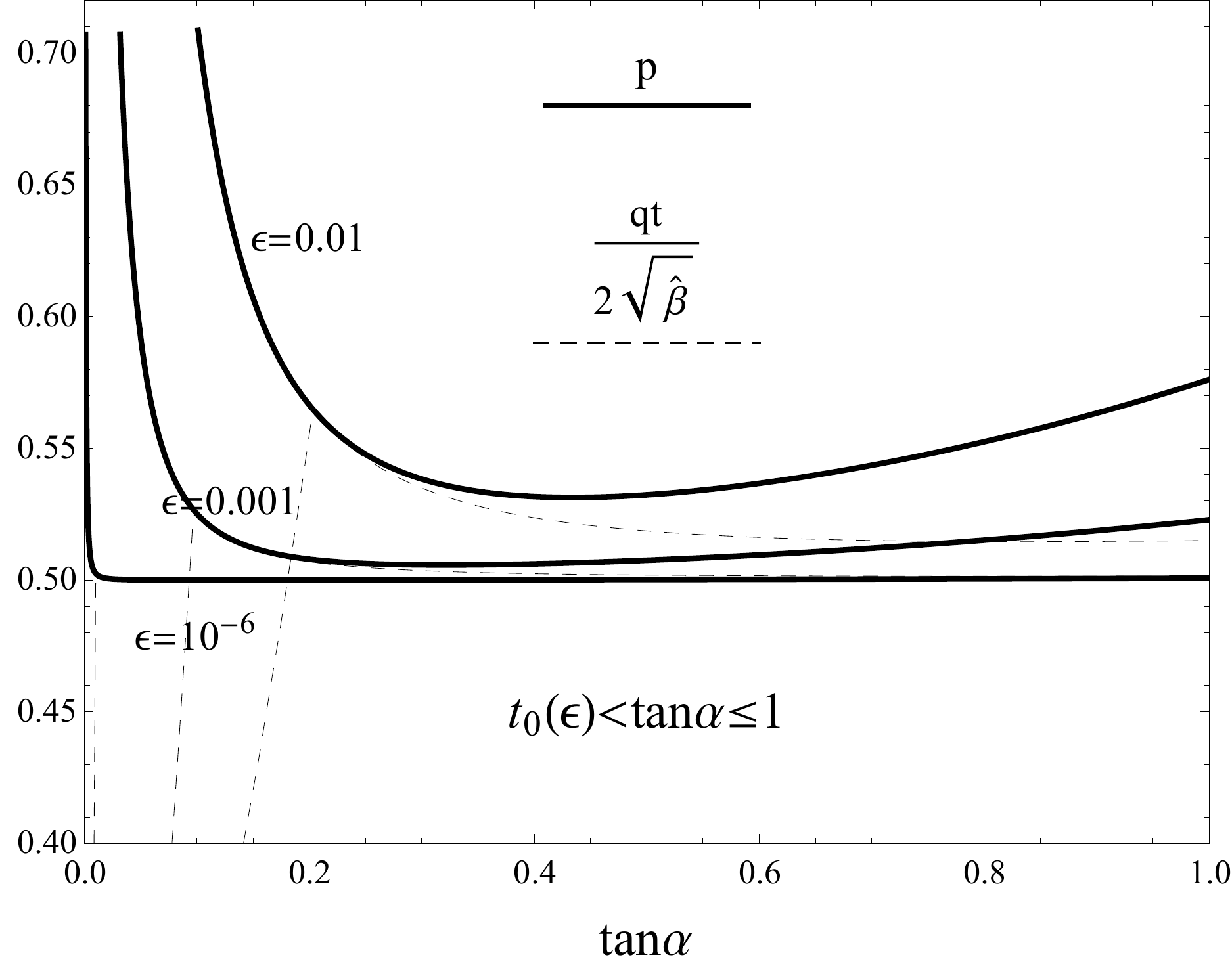}\vspace{0 cm}\caption{ $ p, \frac{q\,t}{ 2\sqrt{\hat \beta }}$  plotted versus $\tan\alpha$ for different values of the material parameter $\epsilon$}
\label{figc}
\end{figure}

In Fig.\ref{figc} the quantities 
$ p, \frac{q\,t}{2\sqrt{\beta}}$ are  plotted in the range $t_0(\epsilon)<\tan\alpha\leq 1$, for different values of the material parameter $\epsilon$. Notice that 
$p<1$ for all the values of $\varepsilon$ here considered.
Then \eqref{limmu} and \eqref{a31} entail to conclude $\widehat\mu_\infty= 0$. 
This shows that in this limit case the optimal tensegrity is `infinitely more efficient' than the \textit{continuum} optimal column having the same span and carrying the same load.
It is worth noticing that in the special case $\bar q=1$ the inequalities \eqref{a31} and \eqref{a32} are strictly satisfied (the two terms are equal). Thus  we may conclude that,  in the limit case of very small load or very large span, for $\bar q\neq 1$ the mass of the optimal tensegrity is  given only by the struts,  whereas for $\bar q=1$ the masses of struts and cables are  equal. 

\begin{figure}
\centering 
\includegraphics[height=5.5cm]{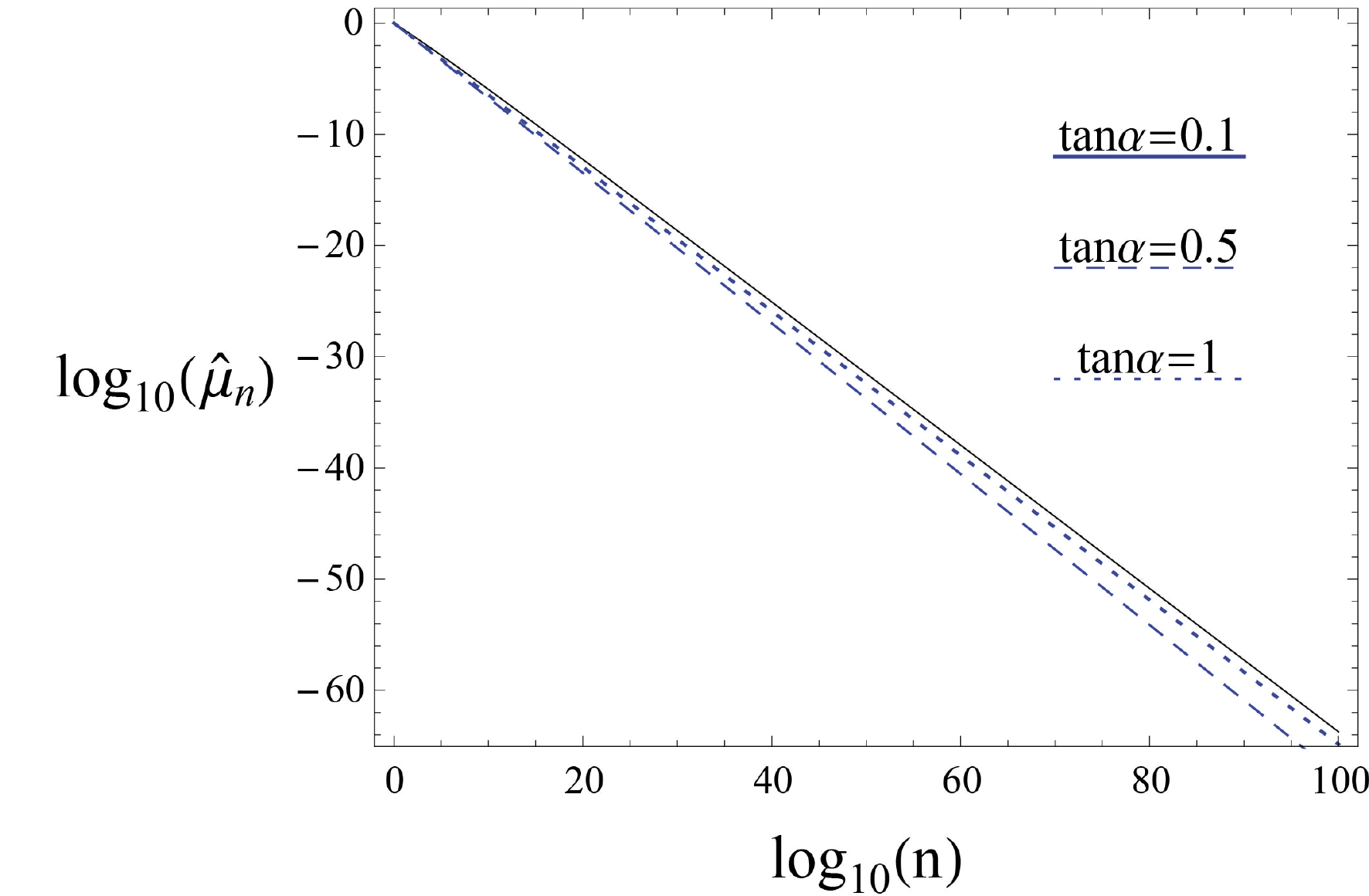}\vspace{0 cm}\caption{ $\widehat \mu_{n}$  plotted versus the complexity $n$  for different values of the geometrical parameter 
$\tan\alpha$.  Here we assumed $\epsilon=0.001$.}
\label{figd}
\end{figure}

In Fig.\ref{figd} the \textit{optimal} relative mass $\widehat \mu_{n}$ is plotted versus the complexity $n$  for different values of $\tan\alpha$. The figure shows the power law relation with the relative mass decreasing as the complexity increases.

\subsection{Limit dimension}
We  continue our analysis by showing that in the limit case of infinite complexity   the   
\textit{optimal} tensegrity has a fractal dimension $D<3$.
 To this end the tensegrity is considered as the union of  cylindrical subsets of $\mathbb{R}^3$ occupying, at every complexity $n\in \mathbb N$, the region $\Omega_n\subset\mathbb{R}^3$. Following \cite{Falconer} (Section 3.1), we recall that the (box-counting) dimension  of any set $\Omega$ is given by
\begin{equation}
 {\rm dim}_B\,\Omega=\displaystyle\lim_{\delta \to 0} \dfrac{\log(N(\delta))}{-\log(\delta)} .
\label{dim}
\end{equation}
 Here $N(\delta)$ is the number of elements in  the smallest set of cubes of side $\delta$ that covers the set of both bars and cables.	
 
In order to apply \eqref{dim}, first we show  that, at any order of complexity, all  bars and  cables are slender cylinders, {\it e.g}. the diameters of their cross sections are always much smaller than their lengths. To this end we note that, since all sections have the same shape, the diameter $\delta_{n,i}$ of a generic section can be written as $\delta_{n,i}=\gamma\sqrt{A_{n,i}}$, where $\gamma$ depends on the shape of the section. 
Then,   for a generic bar  we can write (see \eqref{phie})
\begin{equation}
\left( \dfrac{\delta_{n,i}}{l_{n,i}}\right) ^2=\gamma^2\dfrac{A_{n,i}}{l_{n,i}^2}=\gamma^2\xi \sqrt{\frac{N_{n,i}}{ E\,l_{n,i}^2}}=\xi \sqrt{\phi_{n,i}}\ll 1. \label{ggg}
\end{equation}
A similar equation can be written for the cables.

Consider now separately the sections of the bars and of the cables in order to  determine the smallest cross section and fix $\delta$ as a submultiple of its diameter
\begin{equation}
\delta  =  \min\{\delta_s, \delta_c\},
\label{del}
\end{equation}
where
\begin{equation}
\delta_s=\frac{\gamma}{m}\sqrt{ \min_{ i=0,1,...,n}\{A^s_{n,i}}\}
\label{delb}
\end{equation}
\begin{equation}
\delta_c=\frac{\gamma}{m}\sqrt{ \min_{\substack{j=1,....,n\\ k=0,1,....,j-1}}\{A^c_{j,k}}\},
\label{dels}
\end{equation}
for    a  large enough integer $m$. Here  $A^s$ are the areas of struts section, whereas $A^c$ are the areas of the cables sections.
Thus, by \eqref{pchi}  
\begin{equation}
\begin{array}{lll}
A^s_{n,i}&=&\xi l_{n,i}  \sqrt{\dfrac{N_{n,i}}{E}}= \xi  L a_{n,i} \sqrt{\dfrac{b_{n,i}P_f^{(n)}}{E}}\vspace{0.15cm}\\
 & =&\dfrac{\epsilon_y \xi L}{2^n}\left( \frac{\bo}{(\bo-1)\tan^3\alpha}\right) ^{i/2}\left( \sqrt{(\bo-1) \tan^{3}\alpha}  \right)^n \left(\dfrac{q}{2\sqrt{\bo}}\right) ^n ,      
\end{array}
\end{equation}
where, $P_f^{(n)}$ is the limit load at the given complexity, obtained by the inversion of \eqref{chif00}

\begin{equation}P_f^{(n)}=E L^2  \xi ^2 \epsilon ^2 \left(2\frac{\sqrt{\bo
   }}{q}\right)^{-2 n}.\label{Pfn}\end{equation}
Since, by \eqref{tlim} and \eqref{blim}, $\frac{\bo}{(\bo-1)\tan^3\alpha}>1$, the minimum area of the struts is $A^s_{\min}=A^s_{n,0}$.
Then, from \eqref{qqq} and \eqref{delb}
\begin{equation}
\begin{array}{ll}
\delta_s&=\frac{\gamma}{m}\sqrt{\epsilon_y }\xi L \left(  \frac{q \sqrt{ (\bo-1)\tan^3\alpha}}{4\sqrt{ \bo }}  \right)^{n/2}=\frac{\gamma}{m}\sqrt{\epsilon_y }\xi L \left( 
       \frac{\tan\alpha}{2 }\sqrt{ \frac{q}{\bar{q}}} \right)^{n}.
\end{array}
\label{delb1}
\end{equation}

Then, in view of \eqref{cccc}, \eqref{dddd} and \eqref{pchi}, we can write
\begin{equation}
A^c_{j,k}      =\dfrac{ T_{j,k}}{\sigma_y}= d_{j,k}\dfrac{P_f^{(n)}}{E\epsilon_y}\dfrac{1}{2E\epsilon_y \sin\alpha} \left( \frac{\bo}{(\bo-1)\tan\alpha}\right) ^k \left( (\bo-1)\tan\alpha \right)^j P_f^{(n)},     
\end{equation} 
where, by \eqref{tlim} and \eqref{blim}, $ \frac{\bo}{(\bo-1)\tan\alpha}>1$ and $ (\bo-1)\tan\alpha<1$. Then we have that the minimum area of the cables is $A^c_{\min}=A^c_{n,0}$.
Therefore, in view of  \eqref{Pfn}, we have
\begin{equation}\begin{array}{lll}
  \delta_c&=& \dfrac{\gamma \xi \sqrt{\epsilon_y }L}{m\sqrt{2\sin\alpha}} \left(  \dfrac{q^2}{4 \bo} (\bo -1)\tan\alpha \right) ^{n/2} \dfrac{\gamma\xi \sqrt{\epsilon_y }L}{m\sqrt{2\sin\alpha}}   \left(  \dfrac{q }{ \bar{q}}\dfrac{\tan\alpha}{2} \right)^{n}.
  \end{array}
\label{dels1}
\end{equation}
Finally, by \eqref{del}, \eqref{delb1} and \eqref{dels1}, we obtain
\begin{equation}
\delta  = \frac{\gamma}{m}\sqrt{\epsilon_y }\xi L \left(  \frac{\tan\alpha}{2} \right)^{n}\left(  \frac{q}{\bar{q}}\right) ^{n/2}\min\left\lbrace 1,\dfrac{1}{\sqrt{2\sin\alpha}}\left( \frac{q}{\bar{q}}\right) ^{n/2}\right\rbrace.  \label{delf}
\end{equation}

Now, in order to apply the definition  \eqref{dim}, we  determine by \eqref{delf} the complexity $n_\delta $ such that the diameter of the covering  is equal to $\delta$:
\begin{equation}
n_\delta =\left\lbrace \begin{array}{lll}
&\dfrac{\log \delta- \log \left(\dfrac{\gamma}{m}\sqrt{\epsilon_y }\xi L\right)   } {\log\left( \dfrac{\tan\alpha}{2 }\left(  \frac{q}{\bar{q}}\right) ^{1/2}\right) }\quad &\mbox{if}\quad\bar{q}<1, \vspace{0.2 cm}\\
&\dfrac{\log \delta -\log \left(\frac{\gamma L \xi \sqrt{\epsilon_y  } }{m\sqrt{2\sin\alpha}}\right)}{\log \left( \dfrac{\tan\alpha}{2} \left(  \frac{q}{\bar{q}}\right) \right)}\quad &\mbox{if}\quad\bar{q}\geq 1.
 \end{array}\right . 
\label{ndel}
\end{equation}
Further, we find that the volume $V(\delta)$ of the tensegrity at the complexity $n_\delta$ under the load  $P^{(n_\delta)}_ f $ (see \eqref{Pfn}) is
\begin{equation} \begin{array}{lll}
V(\delta)&=&\dfrac{1}{\rho}m_{b}(L, P_f^{(n_\delta)})\mu_{n_\delta}\\
&=&\epsilon_y \xi^2 L^3 \left(  \frac{q\,}{ 2\sqrt{ \bo } } \right)^{n_\delta}\left( p^{n_\delta}+  \left(  \frac{q\,}{ 2\sqrt{ \bo } } \right)^{n_\delta} (t^{n_\delta} -1) \right) \vspace{0.15cm}\\

&=&  \epsilon_y \xi^2 L^3\left(  \frac{q\,p}{ 2\sqrt{ \bo } } \right)^{n_\delta} \left[1 +  \left(  \frac{q\,t}{ 2p\sqrt{ \bo } } \right)^{n_\delta}  -  \left(  \frac{q}{ 2p\sqrt{ \bo } } \right)^{n_\delta}\right] .
\end{array}
\label{V} \end{equation}
To evaluate   the limit of $\log( V(\delta) )$ as $\delta\rightarrow 0$,  we use \eqref{a31}$_1$, so that,
since $t>1$, we get 
\begin{equation}
\begin{array}{lll}
\lim_{\delta \to 0} \log( V(\delta) )& =&\dfrac{  \log(  \frac{q\,p}{ 2\sqrt{ \bo } } ) \log( \delta ) }{{\log \left( \dfrac{\tan\alpha}{2} \left(  \dfrac{q}{\bar{q}} \right) ^\eta \right)}}\vspace{0.15cm}\\
& =&\dfrac{  \log \left( \frac{q}{ 4 }(1+\frac{\tan^3\alpha}{\bar{q}}) \right) \log( \delta )}{{\log \left( \dfrac{\tan\alpha}{2} \left(  \dfrac{q}{\bar{q}} \right) ^\eta \right)}}\,,
\end{array} 
\label{logV}
\end{equation} 
where 
\begin{equation}
\eta=\left\lbrace \begin{array}{lll}
1&\mbox{if}& \quad  \bar{q}\ge1\vspace{0.15cm}\\
\frac{1}{2}&\mbox{if}&\quad   \bar{q}<1.
\end{array}\right. 
\end{equation}
For small values of $\delta$, the number of cubes in the set covering all the cylindrical  members of the tensegrity can be  evaluated as (see \cite{Falconer})                  
\begin{equation}
N(\delta)=\dfrac{V(\delta)}{\delta^3},
\label{N} 
\end{equation}            
so that, by \eqref{dim},  we have
\begin{equation}
\begin{array}{lll}
D&=& 3  - \lim_{\delta \to 0}\dfrac{ \log (V(\delta))}{\log(\delta)}\vspace{0.15cm}
 =3  - \dfrac{  \log \left( \frac{q}{ 4 }(1+\frac{\tan^3\alpha}{\bar{q}}) \right)}{{\log \left( \dfrac{\tan\alpha}{2} \left(  \frac{q}{\bar{q}} \right) ^\eta \right)}}\vspace{0.15cm}\\
 &=&\dfrac{  \log \left( 2 (1+\frac{\bar{q}}{\tan^3\alpha})\left(  \frac{\bar{q}}{q} \right) ^{3\eta-1} \right)}{{\log \left( \frac{2}{\tan\alpha} \left(  \frac{\bar{q}}{q} \right) ^\eta \right)}}\,.
\end{array}
\label{dimA!} 
\end{equation}    

In Fig.\ref{Fig10}  we plot the fractal dimension  $D$ versus $\varepsilon$ and $\tan\alpha$. 
Notice that $D<3$ for all the values of the parameters $\tan \alpha$ and $\epsilon$.

 \section*{Appendix: global stability}
\label{stab}
 In this section we study the global stability problem for the typical tensegrity of  complexity $n$ based on an iterative approach. To this end we first show that the global stability analysis of the $n$-th complexity tensegrity can be reduced to the analysis of an order one tensegrity (T-bar) with generic length $l$ and generic applied force $N$. This extends the approach considered in \cite{DMPT} to the case in the paper when all struts are refined. \vspace{0.3 cm}
 
\begin{figure}\vspace{-3.5 cm}
\centering 
\includegraphics[scale=0.4]{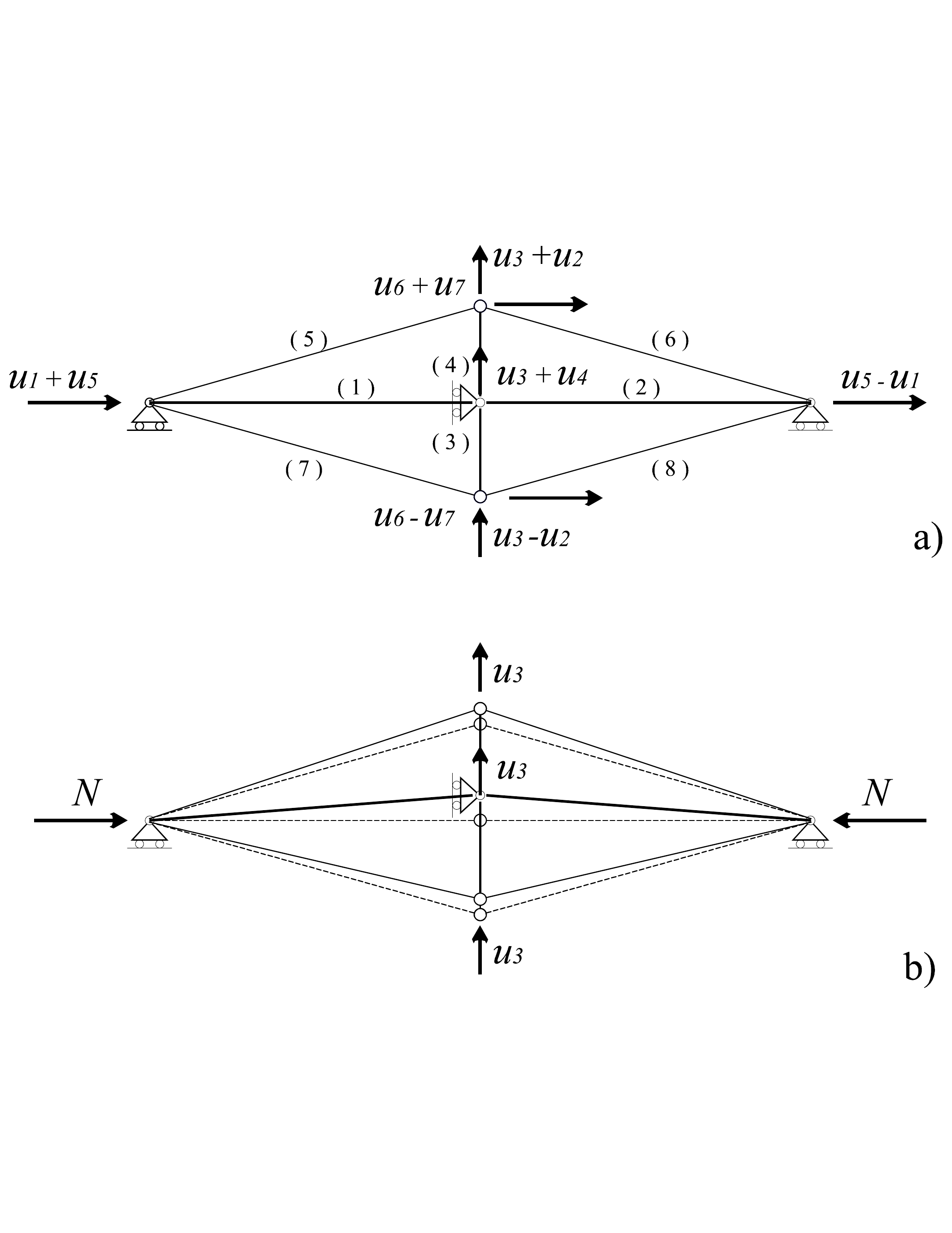}\vspace{-2.2 cm}\caption{Scheme of the global stability. a) Lagrangian variables, b) critical mode. 
}
\label{figb}
\end{figure}

 To this end consider a T-bar of length $l$ undergoing a force $N$.
 We choose as Lagrangian parameters the \textit{generalized} node displacements   $u_i$ (see  Fig.\ref{figb}a), measuring the incremental displacements from the prestressed, loaded configuration. By considering the symmetry properties of the system, these variables are chosen symmetric or antisymmetric with respect to both the vertical axes and the horizontal axes.

The total potential energy (set equal to zero in the prestressed loaded state) can  be written as
\begin{equation}V(\uz , N)=\sum_{k=1}^8 \left [N^{(k)} \Delta l^{(k)} + \frac{E A^{(k)}}{2} 
 \left (\frac{\Delta l^{(k)}}{l^{(k)}} \right )^2  l^{(k)} \right ]
 - 2N u_1,\label{eq1}\end{equation}
 where
 $$\Delta l^{(k)}=||\Delta \xz ^{(k)} + \Tz^{(k)} \uz||- l^{(k)}.$$
 Here $l^{(k)}$ are the lengths of the members with area $A^{(k)}$ and $\Delta l^{(k)}$ are their elongations, whereas $\Delta \xz ^{(k)}$ and $\Tz^{(k)} \uz$ are the axial vector  in the reference configuration and the relative incremental displacement vector between the end joints of the $(k)$-th bar, respectively. 
 
 Our analysis is based on the following assumptions, typical of equilibrium bifurcation analysis for slender structures: \vspace{0.2 cm}

\noindent $-$  we identify the deformed lengths of the bars with the natural ones; \vspace{0.2 cm}

\noindent  $-$   according with the {\it maximum delay convention} (see \cite{PT}) that  the system stays in a (metastable) equilibrium configuration  until it disappears.   

Under these assumption we study the positiveness of the Hessian  (tangent stiffness) matrix
\begin{equation}
\Kz=	\frac{\partial^2 V} {\partial \uz \partial \uz}\mbox{\Huge $\arrowvert$}_{\mbox{\tiny $\uz={\mathbf 0}$ }}  
\label{eq2}
\end{equation}
which, due to our choice of the Lagrangian variables,  is a block diagonal matrix
\begin{equation}\Kz = \left [ \begin{array}{cc} \begin{array}{cc} \Kz_{ss} & \\
 & \Kz_{as} \end{array} & \Oz \\
 \Oz & 
 \begin{array}{cc} \Kz_{sa} & \\
 &K_{aa} \end{array} \end{array} \right ].\label{eq3}\end{equation}
Here $\Kz_{ss}$, $\Kz_{as}$, and $\Kz_{sa}$ are $2 \times 2$ square matrices whereas 
 $K_{aa}$ is a scalar.  The double index notation indicates the symmetric (s) and antisymmetric (a) properties  (see Fig.\ref{figb}), with the first index referring to the horizontal axis  and the second index to the vertical axis.  In particular, $u_1$ and $u_2$ are {\it s-s} variables, $u_3$ and $u_4$ are {\it a-s} variables, $u_5$ and $u_6$ are {\it s-a} variables and  $u_7$ is an {\it a-a} variable. 
Finally, the sub-matrices  have the following expressions:
 \begingroup\makeatletter\def\f@size{9}\check@mathfonts 
 \begin{equation}
 \begin{array}{l}\displaystyle \Kz_{ss}=2\left[
\begin{array}{cc}
 (k_{es}-k_{gs}) \cos 2\alpha+k_{eh}+k_{es}+k_{gs} & (k_{gs}-k_{es}) \sin
   2\alpha   \vspace{0.3 cm} \\
 (k_{gs}-k_{es})  \sin 2\alpha  & (k_{gs}-k_{es}) \cos 2\alpha+k_{ev}+k_{es}+k_{gs}
\end{array}
\right]\vspace{0.3 cm} \\
 \displaystyle\Kz_{as}=2 \left[
\begin{array}{cc}
  (k_{gs}-k_{es}) \cos 2\alpha+k_{gh}+k_{es}+k_{gs}&k_{gh} \\
 k_{gh} &k_{ev}+k_{gh}
\end{array}
\right] \vspace{0.3 cm} \\ 
\displaystyle \Kz_{sa}=2 \left[
\begin{array}{cc}
  (k_{es}-k_{gs}) \cos 2\alpha+k_{eh}+k_{es}+k_{gs}  & (k_{es}-k_{gs}) \cos 2\alpha +k_{gs}+k_{es} \\
 (k_{es}-k_{gs}) \cos 2\alpha +k_{gs}+k_{es} &  (k_{es}-k_{gs}) \cos 2\alpha+k_{gv}+k_{es}+k_{gs}
\end{array}
	\right] \vspace{0.3 cm}
 \\K_{aa}=2((k_{es}-k_{gs}) \cos 2\alpha+k_{gv}+k_{es}+k_{gs}).\end{array}\vspace{0.3 cm}
 \label{ak0}\end{equation}
\endgroup

In order to obtain  general results  valid for a  tensegrity of any complexity $m$, in the present analysis we assume that  the horizontal bars   are \textit{equivalent struts},  having the same axial elastic stiffness of  the internal horizontal tensegrities of complexity $m-1$.  In  the particular case of a tensegrity of order  $m=1$, $k_{eh}$  and $k_{gh}$ are the elastic axial stiffness  and the geometrical stiffness of the \textit{real} horizontal struts
\begin{equation}
    \begin{aligned}
     k_{eh} &=\frac{E A^{(k)}}{l^{(k)}}=\displaystyle  \, \xi  \sqrt{\beta\, E \,N},  \qquad k=1,2,\\
     k_{gh}&=\frac{N^{(k)}}{l^{(k)}}= -\displaystyle\frac {2\beta N\, } {l},
    \end{aligned}   
 \label{ak1}\end{equation} 
where $A^{(1)}=A^{(2)}=\xi\frac{l}{2} \sqrt{\frac{\beta N}{ E}}$.
In the general case $m>1$   $k_{eh}$  and $k_{gh}$ are the elastic axial stiffness  and the geometrical stiffness of the two horizontal tensegrities of complexity $m-1$, having  length $l/2$ and carrying the axial forces $\beta N$. 

Moreover, to avoid complications related to the  evaluation of the stiffness $k_{eh}$ for any internal complexity $m-1$,  we determine a lower bound $\bar{k}_{eh}$ to express $k_{eh}$ as $k_{eh}=\delta_h \bar{k}_{eh}$, where  $\delta_h>1$. 
To this end,  we notice that in the two horizontal  tensegrities of order $m-1$ the axial tension  $\sigma$  in all struts is always lower or equal than $\sigma_y$. Then a lower bound  $\bar{k}_{eh}$ can be determined by  considering  only the $2^{m-1}$ horizontal bars of lengths $\frac{l}{2^m}$, carrying the axial forces $\beta^{m}N$ and by assigning to their  cross sections the lowest feasible  areas  $\frac{\beta^mN}{\sigma_y }$:
\begin{equation}
\bar{k}_{eh} =\dfrac{E}{(l/2)}\Bigl ( \dfrac{\beta^mN}{\sigma_y }\Bigr ) =\dfrac{2\beta^m N}{\epsilon_y l }.
\label{kbarh}
\end{equation} 
The geometrical stiffness of the two horizontal tensegrities of complexity $m-1$ is always given by \eqref{ak1}. 

Similarly to the previous case of horizontal struts, in  the particular case of a tensegrity of order  $m=1$, in  \eqref{ak0}   $k_{ev}$  and $k_{gv}$ are the elastic axial stiffness  and the geometrical stiffness of the \textit{real} vertical struts and we have
\begin{equation}
    \begin{aligned}
     k_{ev} &=\frac{E A^{(k)}}{l^{(k)}}=\displaystyle \xi\sqrt {(\beta - 1)E \, N \tan \alpha},  \qquad k=3,4,\\
     k_{gv}&=\frac{N^{(k)}}{l^{(k)}} = -\displaystyle\frac{2 (\beta - 1)N}{l},
    \end{aligned}   
 \label{ak2}\end{equation} 
where $A^{(3)}=A^{(4)}=\xi\frac{l}{2\tan\alpha} \sqrt{\frac{(\beta-1) N\tan\alpha}{E}}$. In the general case $k_{ev}$  is the axial elastic stiffness of the two internal vertical tensegrities of complexity $m-1$, having lengths $\frac{l\ta}{2}$ and carrying the forces $(\beta-1)N\ta$. As for the horizontal struts,  since it is difficult to find $k_{ev}$ for any complexity $m$,  we will determine a lower bound $\bar{k}_{ev}$  to write $k_{ev}$ as $k_{ev}=\delta_v \bar{k}_{ev}$, with $\delta_v>1$.  Again, since the axial tension  $\sigma$  in all struts is always lower or equal than $\sigma_y$, the lower bound  $\bar{k}_{ev}$ can be determined by  considering  only the $2^{m-1}$ vertical bars of lengths $\frac{l\tan\alpha}{2^m}$, carrying the axial forces $(\beta-1)\beta^{m-1}N\ta$, and by assigning to their  cross sections the lowest feasible  area  $\frac{(\beta-1)\beta^{m-1}N}{\sigma_y }$. Then we can write
\begin{equation}
k_{ev}=\delta_v \dfrac{2E}{l\ta}\dfrac{\beta^{m-1}(\beta-1) N\ta}{\sigma_y  }=\delta_v \dfrac{2\beta^{m-1}(\beta-1) N}{\epsilon_y l }.
\label{kbarv}
\end{equation}
The geometrical stiffness of the two vertical tensegrities of complexity $m-1$ is always given by \eqref{ak2}.
 
Finally, in \eqref{ak0} we have denoted by $k_{es}$ and $k_{gs}$ the elastic stiffness and the geometric stiffness  of the cables: 
  \begin{equation}
    \begin{aligned}
     k_{es} &=\frac{E A^{(k)}}{l^{(k)}}=\frac {E(\beta - 1) \, N} { \sigma_y l}\qquad k=5,6,7,8,\\
     k_{gs}&=\frac{N^{(k)}}{l^{(k)}}=\displaystyle\frac{ (\beta - 1)N}{l},
    \end{aligned}     
  \label{ak4} \end{equation} 
 where 
  \[A^{(k)}=\frac {(\beta - 1) \, N} { 2\sigma_y\cos\alpha}\qquad k=5,6,7,8\]
is the cross section area corresponding to  material failure. Notice that since we assume that all cables are prestressed no slackness effects can occur.

In order to analyze the positivness of the tangent stiffness matrix, we begin by noting that, in view of  equations \eqref{ak0} - \eqref{ak4}, $\displaystyle \Kz_{ss}$ can be written as 
\begingroup\makeatletter\def\f@size{8}\check@mathfonts
\[\begin{array}{ll}
\displaystyle &\Kz_{ss}=\dfrac{2}{ l \left(\tan ^2\alpha+1\right) \epsilon_y}\vspace{0.2cm}\\
&\times\left( \begin{array}{cc}
\begin{array}{cc}
 &\epsilon_y    k_{eh}   l\left(\tan ^2\alpha+1\right)\\
 &+2 N (\beta -1) \left(\epsilon_y   \tan ^2\alpha+1\right) \end{array} &
   -2N (\beta -1) (1-\epsilon_y  ) \tan \alpha \vspace{0.15cm}\\
  -2N (\beta -1) (1-\epsilon_y  ) \tan \alpha
   & \begin{array}{cc}
   &2N (\beta -1)  \left(\tan ^2\alpha+\epsilon_y  \right)\\
   &+ \epsilon_y  k_{ev}  l\left(\tan ^2\alpha+1\right)\end{array} 
\end{array}
\right).
\end{array}\label{kss}\]
\endgroup
It can be found that the two invariants ${\rm det}(\Kz_{ss})$ and ${\rm tr}(\Kz_{ss})$ are always  positive, so that also  the two  eigenvalues $\lambda_{ss1}$ and $\lambda_{ss2}$ associated to this submatrix are  always positive. 
Now we pass to consider the sub matrix $\Kz_{sa}$.
Differently from the previous case, here the positiveness of $\displaystyle \Kz_{sa}$  depends on  the elastic stiffness $k_{eh}$. In view of  equations \eqref{ak0} - \eqref{kbarh},  we  find

\begingroup\makeatletter\def\f@size{9}\check@mathfonts
\[ \begin{array}{lll}
\mbox{\rm tr}(\Kz_{sa})&=&\dfrac{4P}{  \epsilon_y   l\delta_h \left(\tan ^2\alpha+1\right)}   \left(\tan ^2 \alpha+1\right) \Bigl (\:\left(\tan ^2\alpha +1 \right) \beta ^m+(\beta -1)   \left(\epsilon_y  \tan ^2\alpha +1\right)  \vspace{0.2cm}\\
 &+& \epsilon_y  (\beta -1) \left(\:1+\epsilon_y \frac{\tan ^2(\alpha )-1}{2}   \right) \Bigr )
\end{array}
\]
\[ \begin{array}{ll}
\displaystyle &\det(\Kz_{sa})=\dfrac{4N}{  \epsilon_y   l \left(\tan ^2\alpha+1\right)}\times\vspace{0.2cm}\\
& \left |
\begin{array}{cc}
    \delta_h\left(\tan ^2(\alpha )+1\right)\left(\:\left(\tan ^2(\alpha )+1\right) \beta ^m+(\beta -1)
   \left(\epsilon_y   \tan ^2(\alpha )+1\right)\:\right)
   & - (\beta -1) \left(\epsilon_y  
   \tan ^2(\alpha )+1\right)\vspace{0.15cm} \\
 -  (\beta -1) \left(\epsilon_y   \tan ^2(\alpha )+1\right)  & 
   \epsilon_y (\beta -1)  \left(\:1+\epsilon_y \frac{\tan ^2(\alpha )-1}{2}   \right )
\end{array}
\right |.
\end{array}
\]
\endgroup
Now, it can be verified that, for small values of $\epsilon$ (say for $0<\epsilon\leq 1)$, also in this case the two invariants $\det(\Kz_{sa})$ and $\mbox{tr}(\Kz_{sa})$ are  always positive, as  the two eigenvalues $\lambda_{sa1}$ and $\lambda_{sa2}$. Finally, it is also straightforward to verify that  the eigenvalue
$$
\lambda_{aa1}=K_{aa}=\dfrac{2N (\beta -1) \left( \epsilon_y  \left(\sqrt{\tan ^2\alpha+1}-2\right)+2\right)}{\epsilon_y  l \sqrt{\tan ^2(\alpha )+1}}
$$ 
is always positive, for small values of $\epsilon_y $, say  $(0<\epsilon\leq 1)$.

Differently from the previous cases,  the submatrix $\Kz_{as}$  is not always definite positive. Then, in order to find the condition under which the equilibrium is stable, we introduce the  parameter 
\begin{equation}
\phi=\frac{N}{E l^2}.
\label{phie}
\end{equation}
As shown in the following  (Eq. \eqref{ip2}) in all cases here considered  the above parameter  is always  small $(\phi\ll 1)$.  Then,  by \eqref{ak0} - \eqref{phie}, we can write
\begingroup\makeatletter\def\f@size{12}\check@mathfonts
\begin{equation} 
\Kz_{as}=4E l \sqrt{\phi}\left(
\begin{array}{cc}
 \frac{   \beta(1-\epsilon_y ) \tan ^2\alpha  -\tan ^2\alpha-\epsilon_y  }{\epsilon_y  (\tan ^2(\alpha )+1)}\sqrt{\phi} & -\beta \sqrt{\phi} \vspace{0.15cm}\\
 -\beta  \sqrt{\phi}  &  \dfrac{k_{ev}}{2E l \sqrt{\phi}}-\beta \sqrt{\phi} 
\end{array}
\right).
\label{ak7}
\end{equation}
\endgroup
In view of  the  smallness of $\phi$,  the two eigenvalues of $\Kz_{as} $ can be written as
 \begin{equation}\left\lbrace 
 \begin{array}{lll}
\lambda_{as1}&=&4El\dfrac{ \left(  \beta(1-\epsilon_y ) \tan ^2\alpha  -\tan ^2\alpha-\epsilon_y 
   \right)}{\left(t^2+1\right) \epsilon_y  }\phi+ o(\phi^{\frac{3}{2}})\vspace{0.15cm}\\ 
 \lambda_{as2}&=&\left( 2k_{ev}-4El\beta  \phi\right)  + o(\phi^{\frac{3}{2}})
  \end{array}\right. .
\label{eig1}
 \end{equation}

We can note that $\lambda_{as1}$  is positive iff 
 \begin{equation}
  \beta>\widehat{\beta}:=\frac{\epsilon_y +\tan^2 \alpha}{\tan^2 \alpha
   \left(1-\epsilon_y \right)}.
 \label{bet}\end{equation}
To study the positiveness of the second eigenvalue, we note that, in view of \eqref{bet}, we can write  
\begin{equation}\begin{array}{lll}
\lambda_{as2}&=&4El\beta\left( \delta_v \dfrac{\beta^{m-2}(\beta-1) }{\epsilon }-  1\right)\phi   + o(\phi^{\frac{3}{2}})\\
&\geq&4El\bo \left( \delta_v\dfrac{\bo^{m-2}(\bo-1) }{\epsilon_y  }-  1\right)\phi  + o(\phi^{\frac{3}{2}}).
  \end{array}
\end{equation}
So that, we recognized  that   $\lambda_{as2}>0$ for any $m\geq 1$.

In conclusion, we obtain that the lowest  eigenvalue for a generic T-bar  of length $l$ and  subjected to a load $N$, is 
\begin{equation}\lambda_{as1}=\left(\frac{N}{ E l^2}\right) \frac{4E l 
   \left((\beta -1)
   \tan^2 \alpha-\epsilon  \left(\beta 
   \tan^2 \alpha+1\right)\right)}{\epsilon
    \left(\tan^2 \alpha+1\right)} +o\left(\frac{N}{ El^2}\right)
    \label{eig}
    \end{equation}
and it is associated to the  critical mode represented in Fig.\ref{figb}b in which the three joints on the central vertical axis are subjected to the same vertical displacement $u_3$.  Thus, using (\ref{eig}), \textit{we deduce that the T-Bar is stable iff \eqref{bet} holds.}  \vspace{0.3 cm}

The generalization to the case of generic complexity $n$ is based on 
the two following observations. First we remark  that the above stability results do not depend on the elastic axial stiffness of the \textit{equivalent struts}. 
Second we  remark that, while the  {\it generalized stiffness} $\lambda_{as1}$ (with respect the generalized displacement $u_3$) given by \eqref{eig} depends on  the length $l$ and the external load $N$ of the T-Bar,  the stability condition \eqref{bet}  is independent  by these quantities. Thus, to complete our analysis, consider first the order two tensegrity. The stability
of the four T-bars, of length $l_{1,1}$  and $l_{1,0}$ (see Fig.\ref{Fig1}c) subjected to the  forces $ N_{1,1}$ and $ N_{1,0}$ respectively, is ensured by inequality (\ref{bet}). Then, the global stability of the order 2 tensegrity can be analyzed by substituting the four T-bars by 
 {\it equivalent} struts with identical elastic stiffnesses. As shown in the previous subsection also in this case (\ref{bet}) is the stability condition.
  
The stability of the tensegrity of order $n$ is simply obtained by reiterating this approach.  As a consequence, the global stability at any order of complexity is always ensured by the relation \eqref{bet}.

Finally, we  notice that, if  $n\leq c_f$,   the condition  $\phi\ll 1$ is always satisfied at any order of complexity. To this end, we consider the struts $l_{n,i}$ in the tensegrity of complexity $n$. With reference to  these struts,  in view of  \eqref{pchi}, we can write:   
\begin{equation}
\setlength{\jot}{15pt}
\phi_{n,i}=\dfrac{N_{n,i}}{El_{n,i}^2}=\bigl( \xi\epsilon_y\chi( l_{n,i},N_{n,i})\bigr) ^{2}\ll 1\quad \mbox{ for all }      n\geq 1, \quad      i=0,1,....,n .\\          
\label{ip2}
\end{equation}
So that, by \eqref{chil1} the condition \eqref{ip2} is always satisfied.

We conclude  the present analysis by noting that \textit{the condition  $\beta>\widehat{\beta}$ ensures the stability of the tensegrity, whereas $\beta=\widehat{\beta}$  yields to the simultaneous attainment of  critical equilibrium at any order of complexity}.

\end{document}